\documentclass[aip,jcp,reprint,preprintnumbers,amsmath,amssymb,floatfix]{revtex4-1}
\usepackage{graphicx}

\usepackage{amsmath}
\usepackage{mathtools}
\usepackage{amssymb}
\usepackage{amsfonts}  
\usepackage{rotating}
\usepackage{multirow}
\usepackage{dcolumn}

\newcommand{\bra}[1]{\langle #1|}
\newcommand{\ket}[1]{|#1\rangle}

\newcommand{\half}{\frac{1}{2}}

\newcommand{\pieps}{ 4 \pi \varepsilon_0 }

\begin{document}
\edef\marginnotetextwidth{\the\textwidth}

\title{Direct simulation of electron transfer using ring polymer molecular dynamics: Comparison with semiclassical instanton theory and exact quantum methods}

\author{Artur R. Menzeleev}
\author{Nandini Ananth}
\author{Thomas F. Miller, III}
\email{tfm@caltech.edu}

\affiliation{
Division of Chemistry and Chemical Engineering, California Institute of Technology, Pasadena, CA 91125, USA
}

\date{\today}
\begin{abstract}

The use of ring polymer molecular dynamics (RPMD) for the direct
simulation of electron transfer (ET) reaction dynamics is analyzed in the
context of Marcus theory, semiclassical instanton theory, and exact
quantum dynamics approaches.
For both fully atomistic and system-bath representations of condensed-phase ET, we demonstrate that RPMD accurately predicts both ET reaction 
rates and mechanisms throughout the normal and activationless
regimes of the thermodynamic driving force.
Analysis of the ensemble of reactive RPMD trajectories reveals the solvent reorganization mechanism for ET that is anticipated in the Marcus rate theory, and the accuracy of the RPMD rate calculation is understood in terms of its exact description of statistical fluctuations and its formal connection to semiclassical instanton theory for deep-tunneling processes. 
In the inverted regime of the thermodynamic driving force, neither RPMD nor a related 
formulation of semiclassical instanton theory capture the characteristic turnover in the reaction rate; 
comparison with exact quantum dynamics simulations reveals that these 
methods provide inadequate quantization of the real-time electronic-state dynamics in the inverted regime.

\end{abstract}

\maketitle

\section{Introduction}

Condensed-phase electron transfer (ET) reactions are central to 
many biological and synthetic pathways for energy conversion and catalysis.\cite{Huy07,Lew06,Mar85,Gray96_review} 
The development of accurate, robust, and scalable methods for the study of such reactions is thus a key objective in theoretical chemistry. 
Although transition state theories and rate models for ET
have been successfully applied in complex systems,  \cite{Mya05,Jea92,Blu06,Ung99}
methods for the direct simulation and mechanistic study of ET dynamics in general systems remain less fully developed. 
To this end, we explore the use of ring polymer molecular dynamics (RPMD) for the description of prototypical ET reactions between mixed-valence transition metal ions  in water, and we compare the RPMD approach against benchmark semiclassical and quantum dynamics methods.

Fundamental theoretical challenges in the direct simulation of  ET reactions arise due to the coupling of the intrinsically quantum mechanical electronic transitions with slower, classical motions of the surrounding environment.  
Numerous semiclassical and mixed quantum-classical dynamics methods have been developed for the investigation of electronically non-adiabatic reactions reactions, 
\cite{Ehr27,McL64,Tul90,Tul98,Ben00,Mey79,Mey80,Sto97,Cao96,Cao97,nandinitfm} 
but 
existing methods do not enable mechanistic studies that are independent of dividing surface assumptions in general systems;
nor do they yield dynamical trajectories that preserve the equilibrium Boltzmann distribution\cite{Nie01,Sch08} and allow for the use of rare-event sampling methodologies.\cite{Bol02}
New methods are needed to 
accurately describe coupled electronic and nuclear dynamics and 
to enable the efficient and robust simulation of long trajectories that bridge the multiple timescales 
 of ET reactions in complex systems.

RPMD \cite{Cra04} 
is an approximate quantum dynamical method that is based on the imaginary-time path integral formulation of statistical mechanics.\cite{Fey65,Cha81}
It provides an isomorphic classical molecular dynamics model for the real-time evolution of a quantum mechanical system. 
Previous applications of RPMD include studies of molecular liquids, \cite{tfm05rpmdb,tfm05rpmdc,tfm06rpmd,Hab09,Mar11,Cra06} hydrogen transfer rates,\cite{Cra05,Col08,Mar08,Boe11}  and tunneling processes in low-dimensional systems.\cite{Col09,Ric09,Cra05b} 
A key feature of the RPMD method is that it yields real-time molecular dynamics trajectories that  
preserve the exact quantum Boltzmann distribution and exhibit time-reversal symmetry. \cite{Cra04,Par84}
These properties allow RPMD to be used in combination with rare-event sampling methods for the trajectory-based analysis of quantum mechanical tunneling processes in systems involving thousands of atoms. \cite{Boe11,tfm10a}
We have recently extended RPMD to describe electronic and nuclear dynamics, including solvated electron diffusion \cite{tfm08} and non-adiabatic electron injection into liquid water. \cite{tfm10a}

In the current paper, RPMD is used to directly simulate ET dynamics in both atomistic and system-bath representations for mixed-valence
ET in water.
The calculated rates and mechanisms are analyzed in the context of semiclassical and exact quantum methods.
A description of the employed methodologies is provided in Section \ref{sec:methods}, and Section \ref{sec:models} presents the details of the atomistic and system-bath representations. Calculation details are given in Section \ref{sec:calc_details}, and a discussion of the results is presented in  Section \ref{sec:results}.

\section {Methods}
\label{sec:methods}

Several methods are utilized to investigate ET rates and mechanisms, including RPMD, semiclassical instanton theory, exact quantum-mechanical dynamics, and the classical Marcus rate theory for ET. These methods are summarized below.

\subsection{Ring Polymer Molecular Dynamics}

The RPMD equations of motion for a quantized electron and $N$ classical particles 
are\cite{Cra04,tfm08}
\begin{eqnarray}
\dot{\mathbf{v}}^{(\alpha)}&=&\omega_{n}^{2} \left( \mathbf{q}^{(\alpha+1)}+ \mathbf{q}^{(\alpha-1)}-2 \mathbf{q}^{(\alpha)}\right) \nonumber\\
&-&\frac{1}{m_\mathrm{e}}\nabla_{\mathbf{q}^ {(\alpha)}} U_{\mathrm{ext}}\left(\mathbf{q}^{(\alpha)},\mathbf{Q}_{1},\dots,\mathbf{Q}_{N}\right) 
\label{eq:rpmd_eom1}
\end{eqnarray}
and
\begin{equation}
\dot{\mathbf{V}}_j=-{\frac{1}{nM_j}\sum_{\alpha=1}^{n} \nabla_{\mathbf{Q}_{j}} U_{\mathrm{ext}}(\mathbf{q}^{(\alpha)},\mathbf{Q}_{1},\ldots,\mathbf{Q}_{N})},
\label{eq:rpmd_eom2}
\end{equation}
where $\mathbf{q}^{(\alpha)}$ and $\mathbf{v}^{(\alpha)}$ are the position and velocity vectors of $\alpha^{\text{th}}$ ring polymer bead, $\mathbf {Q}_{j}$ and $\mathbf{V}_{j}$ are the position and velocity vectors of the $j^\text{th}$ classical particle, and  $n$ is the number of imaginary-time ring-polymer beads.
The intra-bead harmonic frequency is  $\omega_{n}/(\beta \,\hbar)$, where $\beta$ is the reciprocal temperature. The masses of electron and classical particles are $m_\mathrm{e}$ and $M_j$, respectively, $U_{\mathrm{ext}}(\mathbf{q}^{(\alpha)}, \mathbf{Q}_{1},\ldots,\mathbf{Q}_{N})$ is the potential energy function of the system, and $\mathbf{q}^{(0)}=\mathbf{q}^{(n)}$. Eqs.~\ref{eq:rpmd_eom1}~and~\ref{eq:rpmd_eom2} generate a classical dynamics that we employ as a model for the real-time dynamics of the system.\cite{tfm08} In the limit of large $n$, these dynamics preserve the exact Boltzmann distribution.\cite{Par84}  

As in classical  formulations of the thermal rate constant, \cite{Wig32,Eyr35,Kec60} the RPMD rate can be expressed as\cite{Cra05,Cra05b} 
\begin{equation}
\label{eq:rpmd_rate}
k_{\mathrm{RPMD}}=\lim_{t\to\infty}\kappa(t) k_{\mathrm{TST}}.
\end{equation}
Here, $k_{\mathrm{TST}}$ is the transition state theory (TST) approximation for the rate for a dividing surface $\xi(\mathbf{r})=\xi^\ddagger$, where $\xi(\mathbf{r})$ is a collective variable, $\mathbf{r}=\left\{\mathbf{q}^{(1)},\dots,\mathbf{q}^{(n)}, \mathbf{Q}\right\}$ is the full position vector for the system, 
and $\mathbf{Q}=\{\mathbf {Q}_{1},\dots,\mathbf {Q}_{N}\}$ denotes the set of 
classical particle positions.
The prefactor, $\kappa(t)$, is the time-dependent transmission coefficient that accounts for recrossing of trajectories through the dividing surface. 
An important feature of RPMD is that calculated rates and mechanisms are independent of the choice of TST dividing surface, as in exact quantum and exact classical dynamics. \cite{Cra05,Cra05b,Mil73}

The TST rate in Eq.~\ref{eq:rpmd_rate} is calculated using \cite{Col08,Cha78,Ben77}
\begin{equation}
\label{eq:rpmd_tst}
k_{\mathrm{TST}}= (2 \pi \beta) ^{-1/2} {\left \langle g_\xi \right \rangle_{\mathrm{c}}}\frac{e^{-\beta \Delta F(\xi^\ddagger)}} { \int_{-\infty}^{\xi^\ddagger} d \xi e^{-\beta \Delta  F(\xi)} }.
\end{equation}
Here, $ F(\xi)$ is the free energy (FE) along $\xi$
\begin{equation}
\label{eq:delta_f}
e^{-\beta \Delta F(\xi')}=\frac{\langle \delta (\xi(\mathbf{r})-\xi') \rangle} {\langle \delta (\xi(\mathbf{r})-\xi_{r}) \rangle},
\end{equation}
where $\xi_r$ is a reference point in the reactant region, and \cite{Car89,Sch03,Wat06}
\begin{equation}
\label{eq:f_xi}
g_\xi (\mathbf{r})=\left[ \sum_{i=1}^{d}{\frac{1}{m_i}\left(\frac{\partial \xi(\mathbf{r})}{\partial r_i}\right)^2}\right]^{1/2}.
\end{equation}
The scalar $r_i\in \left\{ \mathbf{r} \right\}$ in this equation indicates either a ring-polymer or classical particle degree of freedom, $m_i$ is the corresponding mass, and $d$ is the total number of degrees of freedom in the system.
In Eqs.~\ref{eq:rpmd_tst}~and~\ref{eq:delta_f}, $\langle \dots \rangle$~denotes the equilibrium ensemble average
\begin {equation}
\langle \dots \rangle= \frac{\int  d \mathbf{r}\int d \mathbf{v}  \; e^{-\beta H_n( \mathbf{r}, \mathbf{v})} (\dots)}{ \int  d \mathbf{r} \int d \mathbf{v}  \; e^{-\beta H_n(\mathbf{r}, \mathbf{v})}},
\end{equation}
and $\langle \dots \rangle_{\mathrm{c}}$ denotes the average in the constrained ensemble
\begin {equation}
\langle \dots \rangle_{\mathrm{c}}= \frac{\int  d \mathbf{r} \int d \mathbf{v} \;e^{-\beta H_n(\mathbf{r}, \mathbf{v}) } (\dots) \delta(\xi(\mathbf{r})-\xi^\ddagger)}{ \int  d \mathbf{r} \int d \mathbf{v}  \; e^{-\beta H_n(\mathbf{r}, \mathbf{v}) } \delta(\xi(\mathbf{r})-\xi^\ddagger) }.
\end{equation}
Here,
\begin{eqnarray}
H_n(\mathbf{r}, \mathbf{v}) =\sum_{j=1}^N \frac {1}{2}M_j \mathbf{V}^2_j+ \sum_{\alpha=1}^n\frac{1}{2} m_\mathrm{b}\left( \mathbf{v}^{(\alpha)}\right)^2+U_n(\mathbf{r}),\;\;\;\;
\end{eqnarray}
where $m_{\mathrm{b}}$ is the fictitious Parrinello-Rahman mass,\cite{Par84}  ${\mathbf{v}=\left\{\mathbf{v}^{(1)},\dots,\mathbf{v}^{(n)}, \mathbf{V}_1,\ldots,  \mathbf{V}_N\right\}}$,  and
\begin{eqnarray}
\label{eq:potential_energy}
U_n(\mathbf{r})&=&\frac{1}{n}\sum_{\alpha=1}^n\frac{1}{2}m_\mathrm{e} \omega_n^2 \left(\mathbf{q}^{(\alpha)}-\mathbf{q}^{(\alpha-1)}\right)^2 \nonumber \\
&+&\frac{1}{n} \sum_{\alpha=1}^n U_{\mathrm{ext}}\left(\mathbf{q}^{(\alpha)},\mathbf{Q}\right)
\end{eqnarray}
is the full potential energy function for the ring polymer.

The transmission coefficient in Eq.~\ref{eq:rpmd_rate} is obtained from the flux-side correlation function using
\begin{equation}
\label{eq:kappa_final}
\kappa(t)=\frac{\left \langle \dot \xi_0 \;h{\left(\xi(\mathbf{r}_t) - \xi^\ddagger\right)} \right \rangle_\mathrm{c} }{\left \langle \dot \xi_0 \; h{\left(\dot \xi_0\right)} \right \rangle_{\mathrm{c}}},
\end{equation}
where $h(\xi)$ is the Heaviside function, $\dot \xi_0$ is the initial velocity of the collective variable in an RPMD trajectory released from the dividing surface, and $\mathbf{r}_t$ is the time-evolved position of the system along that trajectory. \cite{Col08}

\subsection{Semiclassical Instanton Theory}

The ``Im F" premise in semiclassical rate theory relates  the thermal rate constant in the deep-tunneling regime to the analytical continuation of the partition function into the complex plane,\cite{Ben94,Cha75,Cal77,Mil97,Cal83, Han88}
\begin{equation} 
\label{eq:sci_rate}
k \approx \frac{2}{\beta\hbar Q_\textrm{r}}\textrm{Im}\ \!R,
\end{equation}
where $Q_\textrm{r}$ is the reactant partition function and $\textrm{Im}\ R$ is the imaginary part of 
the analytical continuation of the partition function for the full system.
In the steepest-descent limit, the ``Im F" description is equivalent to the flux-side time correlation formulation\cite{Mil75} of semiclassical instanton (SCI) theory.\cite{Alt11}  We adapt this approach to describe the transfer of a single quantized electron in a classical solvent.

The partition function for the full system in the ring-polymer representation can be expressed
\begin{equation}
Q_n=c \int  d \mathbf{Q} \; I_n\!\left(\mathbf{Q}\right)
\end{equation}
where $c=\prod_{j=1}^N\frac{M_j}{2\pi\beta\hbar^2}$,
\begin{equation}
I_n\!\left(\mathbf{Q}\right)=\left(\frac{m_{\textrm{e}}\omega_n}{2\pi\hbar}\right)^{-n/2}
\int  d \{q^{(\alpha)}\} e^{-A(\{q^{(\alpha)}\};\mathbf{Q})/\hbar},
\end{equation}
and 
\begin{equation}
\label{eq:rp_action}
A(\{{q}^{(\alpha)}\};\mathbf{Q})=(\beta\hbar) U_n(\mathbf{r})
\end{equation} 
is the classical action for a periodic trajectory in imaginary time.  
The notation presented here assumes that the quantized electron moves in a single dimension.
At each solvent configuration, the steepest-descent approximation to $I_n\!\left(\mathbf{Q}\right)$ is obtained by expanding $A(\{q^{(\alpha)}\};\mathbf{Q})$ to second order about its global minimum $\{\tilde{q}^{(\alpha)}\}$, for which the electron ring-polymer coordinates obey the stationary condition
\begin{eqnarray}
\label{eq:sci_stationary}
&&
\frac1n\sum_{\alpha=1}^n \Biggl|
\frac{\partial}{\partial q^{(\alpha)}} \left.U_{\mathrm{ext}}\left(q^{(\alpha)},\mathbf{Q}\right) \right|_{q^{(\alpha)}=\tilde{q}^{(\alpha)}}-\Biggr.
\nonumber\\
&& 
\qquad\qquad
\Biggl.\omega_{n}^{2} \left( \tilde{q}^{(\alpha+1)}+ \tilde{q}^{(\alpha-1)}-2 \tilde{q}^{(\alpha)}\right) \Biggr| =0.
\end{eqnarray}
The steepest-descent approximation yields
\begin{equation}
\label{eq:sci_integrand}
I_n\!\left(\mathbf{Q}\right)=\frac{1}{\sqrt{\textrm{det}\ \textbf{K}}}e^{-A(\{\tilde{q}^{(\alpha)}\};\mathbf{Q})/\hbar},
\end{equation}
where $\textbf{K}$ is the Hessian matrix given by 
\begin{equation}
K_{\mu\nu}=\frac{\omega_n}{\hbar}\frac{\partial^2}{\partial q^{(\mu)}\partial q^{(\nu)}}\left.A(\{\tilde{q}^{(\alpha)}\};\mathbf{Q})\right|_{\{q^{(\alpha)}\}=\{\tilde{q}^{(\alpha)}\}},
\end{equation}
and where $(\textrm{det}\ \textbf{K})=\prod_{i=1}^n \eta_i^2$ is obtained from the normal mode frequencies, $\{\eta_i\}$.

For a reaction with a barrier, a saddle point satisfies the stationary condition in Eq.~\ref {eq:sci_stationary}, and 
the Hessian matrix exhibits an imaginary normal-mode frequency, $\eta_1$.  
By analytically continuing $\eta_1$ onto the real axis, and by integrating out the zero-frequency normal mode that is associated with the cyclic permutation of the ring-polymer beads, we obtain the steepest-descent SCI rate\cite{Alt11}
\begin{equation}
\label{eq:sci_rate_sd}
k_{\mathrm{SCI}}=\frac{c}{Q_\textrm{r}}  \int  d \mathbf{Q} \; \mathcal{I}_n\!\left(\mathbf{Q}\right),
\end{equation}
where 
\begin{equation}
\label{eq:ac_sci_integrand}
\mathcal{I}_n\!\left(\mathbf{Q}\right)=\left(\frac{m_e B_n \omega_n^3}{2 \pi \hbar}\right)^{1/2}\frac{1}{\sqrt{\textrm{det}'\ \textbf{K}}}e^{-A(\{\tilde{q}^{(\alpha)}\};\mathbf{Q})/\hbar},
\end{equation}
$(\textrm{det}'\ \textbf{K})=\prod' |\eta_i|^2$ is obtained from a product that excludes the zero-frequency mode, and 
%
\begin{equation}
B_n=\sum_{\alpha=1}^{n} (\tilde{q}^{(\alpha+1)}-\tilde{q}^{(\alpha)})^2.
\end{equation}

Formal connections between path-integral statistics and reactive tunneling have long been recognized. \cite{Cep87, Cha81,Kuk87,Ale88,Alt11,Ric11} In particular, Althorpe and coworkers\cite{Ric09} have recently emphasized the 
connection between the TST limit of RPMD and the reversible action work (RAW) formulation of SCI theory.\cite{Mil97,Mills_Chapter17}  To the extent that Eq.~\ref{eq:sci_rate_sd} is an harmonic approximation to the RAW SCI formulation,\cite{Ric09}
\begin{equation}
k_{\mathrm{RPMD}}=\left( \frac{\kappa_\mathrm{o}}{\alpha} \right)  k_{\mathrm{SCI}},
\label{eq:alpha_correction}
\end{equation}where $\alpha=2\pi(\beta \hbar |\eta_1|)^{-1}$,
and $\kappa_\mathrm{o}$ is the transmission coefficient through a dividing surface that minimizes the recrossing of RPMD trajectories.

\subsection{Exact Quantum Dynamics}
\label{sec:quapi}
We obtain numerically exact quantum dynamics for ET using the Quasi-Adiabatic Path Integral method (QUAPI).\cite{Mak92,Top96,Top93a,Top93b,Mak93} 
The method is applied to a redox system composed of two diabatic electronic states and a coordinate representing polarization of the solvent dipole field; the solvent coordinate is in turn linearly coupled to a harmonic oscillator bath.

The Hamiltonian for the redox system \cite{Top96}
\begin{equation}
\label{eq:tls_hamiltonian}
	{H}_\mathrm{S} =\frac{p_s^2}{2m_{\mathrm{s}}} + 
	\left(\begin{array}{cc}
	   V_{11}(s) & V_{12}(s)\\
	   V_{12}(s) & V_{22}(s)\\
   \end{array}\right)\\,
\end{equation}
where $s$ is the solvent coordinate, $p_s$ is the conjugate momentum, and $m_{\mathrm{s}}$ is the effective solvent  mass. Here, $V_{11}(s)$ and $V_{22}(s)$ are diabatic states corresponding to reactant and product states for ET, and $V_{12}(s)$ is the electronic coupling. The Hamiltonian describing the bath modes and their coupling to the solvent coordinate is
\begin{equation}
   H_\mathrm{B}= \sum_{j=1}^f \frac{P_j^2}{2M_j} + \sum_{j=1}^f\half M_j\omega_j^2
   \left(Q_j-\frac{c_js}{M_j\omega_j^2}\right)^2,
   \label{eq:quapi_bath_hamiltonian}
\end{equation}
where $M_j$, $\omega_j$, and  $Q_j$ are the mass,  frequency and the position of the  $j^\text{th}$ bath mode, respectively, and $c_j$ is the strength of the coupling between the $j^\text{th}$ bath mode and the solvent coordinate. 

The exact quantum mechanical rate constant can be expressed in terms of the symmetrized real-time flux-flux correlation function, \cite{Mil83}
\begin{equation}
	k_\text{Q} = \lim_{t'\to\infty} \frac{1}{Q_{\mathrm{R}}}\int_{0}^{t'} C_{\mathrm{FF}}(t) dt,
	\label{eq:kquant}
\end{equation}
where  
\begin{equation}
	C_{\mathrm{FF}}(t) = \mathrm{Tr}[\mathcal{F}e^{iHt_{\mathrm{c}}^{*}/\hbar}\mathcal{F}e^{-iHt_{\mathrm{c}}/\hbar}], 
	\label{eq:cff_defn}
\end{equation}
and  $Q_{\mathrm{R}}$ is the reactant partition function.
Here, ${H= H_\mathrm{S}+H_\mathrm{B}}$ is the full ET Hamiltonian, ${\mathcal{F}= \frac{i}{\hbar}[ H, \mathcal{P}_{\mathrm{2}}]}$ is the operator for the flux between the reactant and product electronic states, ${\mathcal{P}_\mathrm{2}=\ket{2}\bra{2}}$ 
is the projection operator for  the product electronic state, and ${t_{\mathrm{c}}=t-i\beta\hbar/2}$ is the complex time. 
The propagators are discretized into $\mathcal{N}$ time slices of length $\Delta t_{\mathrm{c}}$, 
and the trace in Eq.~\ref{eq:cff_defn} is expanded to yield
\begin{eqnarray}
\label{eq:cff_expand}
C_{\mathrm{FF}}&&(t)= \int_{-\infty}^{\infty} d \mathbf{Q}_0
 \Bigg \langle \mathbf{Q}_0 \; \Bigg| \;  \bra{s_1, \sigma_1}\mathcal{F}\ket {s_{2\mathcal{N}+2}, \sigma_{2\mathcal{N}+2}} \nonumber \\
&&\times\prod_{k=\mathcal{N}+3}^{2\mathcal{N}+2}\bra{\sigma_k,s_k}e^{iH\Delta t_{\mathrm{c}}^*/\hbar}\ket{\sigma_{k-1},s_{k-1}} \nonumber \\
&&\times \; \bra{s_{\mathcal{N}+2} ,\sigma_{\mathcal{N}+2}}\mathcal{F}\ket {s_{\mathcal{N}+1}, \sigma_{\mathcal{N}+1}} \nonumber \\
&&\times\left.\left.\prod_{k=2}^{\mathcal{N}+1}\bra{\sigma_k,s_k}e^{-iH\Delta t_{\mathrm{c}}/\hbar}\ket{\sigma_{k-1},s_{k-1}}  \; \right | \mathbf{Q}_0 \right \rangle,
\end{eqnarray}
where $\mathbf{Q}_0$ represents the bath degrees of freedom, 
$s_k$ is the solvent coordinate, 
and $\sigma_k$  is the electronic state at complex time slice $k$. 

The propagators in~Eq.~\ref{eq:cff_expand} are factorized using the quasi-adiabatic short-time approximation \cite{Mak92}
\begin{equation}
	e^{-iH\Delta t_{\mathrm{c}}/\hbar}\approx e^{-iH_{\mathrm{B}}\Delta t_{\mathrm{c}}/2\hbar}
	e^{-iH_{\mathrm{S}}\Delta t_{\mathrm{c}}/\hbar}e^{-iH_\mathrm{B}\Delta t_{\mathrm{c}}/2\hbar}.
	\label{eq:split_prop}
\end{equation}
Analytical integration over the bath modes then yields
\begin{eqnarray}
   C_{\mathrm{FF}}(t)&=&\frac{1}{\hbar^2}\text{Re}\;\left[ C_1(2,2,1,1;t_{\mathrm{c}}) - C_2(2,1,2,1;t_{\mathrm{c}}) \right. \nonumber \\
   &+& \left. C_3(1,1,2,2;t_{\mathrm{c}}) - C_4(1,2,1,2;t_{\mathrm{c}})\right],  
   \label{eq:cff_quapi}
\end{eqnarray}
where  
\begin{eqnarray}
       C_i(&&\sigma_1,\sigma_{\mathcal{N}+1},\sigma_{\mathcal{N}+2},
	\sigma_{2\mathcal{N}+2};t_{\mathrm{c}}) \nonumber\\
	=&& \int ds_1\cdots \int ds_\mathcal{N}
	\int ds_{\mathcal{N}+2}\cdots\int ds_{2\mathcal{N}+1}  \\
	\times&& \sum_{\sigma_2=1}^2 \cdots \sum_{\sigma_{\mathcal{N}}=1}^2
	\sum_{\sigma_{\mathcal{N}+3}=1}^2\cdots\sum_{\sigma_{2\mathcal{N}+1}=1}^2 
	I_i(\mathbf{s},\boldsymbol{\sigma};t_\mathrm{c})\nonumber
	\label{eq:cdef}
\end{eqnarray}
and
\begin{equation}
I_i(\mathbf{s},\boldsymbol{\sigma};t_\mathrm{c}) = V_{12}(s_{1})\;V_{12}(s_{\mathcal{N}+2})
K(\mathbf{s},\boldsymbol{{\sigma}};t_\mathrm{c})\mathcal{I}(\mathbf{s}).
\label{integrandQuapi}
\end{equation}
Here, $s_{2\mathcal{N}+2}=s_1$,  $s_{\mathcal{N}+2}=s_{\mathcal{N}+1}$,  and  we have introduced the notation ${\mathbf{s}=\{s_1, \ldots, s_{2\mathcal{N}+2}\}}$  and ${\boldsymbol{\sigma}=\{\sigma_1,\ldots,\sigma_{2\mathcal{N}+2}\}}$.

In Eq.~\ref{integrandQuapi}, the path-integral expression for the complex-time propagators of the system Hamiltonian is given by  
\begin{eqnarray}
\label{eq:k_prod}
K(\mathbf{s},&&\boldsymbol{\sigma};t_{\mathrm{c}})=\prod_{k=\mathcal{N}+3}^{2\mathcal{N}+2}\bra{\sigma_k,s_k}e^{iH_{\mathrm{S}}\Delta t_{\mathrm{c}}^*/\hbar}\ket{\sigma_{k-1},s_{k-1}} \nonumber\\
&&\times\prod_{k=2}^{\mathcal{N}+1}\bra{\sigma_k,s_k}e^{-iH_{\mathrm{S}}\Delta t_{\mathrm{c}}/\hbar}\ket{\sigma_{k-1},s_{k-1}}.
\end{eqnarray}
The matrix elements in~Eq.~\ref{eq:k_prod}
are obtained using the numerically exact expression 
\begin{eqnarray}
	\bra{s_k,\sigma_k}e^{-iH_{\mathrm{S}}\Delta t_{\mathrm{c}}/\hbar}\ket{s_{k-1},\sigma_{k-1}}=&&\nonumber\\
	\sum_{m=1}^{M_0} \phi_m(s_k,\sigma_k)\phi_m^*(s_{k-1},\sigma_{k-1})&&
	e^{-iE_m\Delta t_{\mathrm{c}}/\hbar},
\label{eq:qm_prop}
\end{eqnarray}
where $\phi_m(s,\sigma)$ and $E_m$ are the eigenstates and eigenenergies of $H_{\mathrm{S}}$, respectively, and $M_0$ is the number of eigenstates included in the expansion. 

The discretized form of the non-local influence functional in Eq.~\ref{integrandQuapi}, which accounts for bath-induced electronic transitions in the system, is 
\begin{equation}
	\mathcal{I}(\mathbf{s})=\mathcal{I}_0\;
	\text{exp}\left(-\sum_{k=1}^{2\mathcal{N}+2}\sum_{k'=1}^k 
	B_{kk'}s_k\;s_{k'}\right),
	\label{eq:bathinf}
\end{equation}
where  $\mathcal{I}_0$ is the partition function of the uncoupled bath oscillators. \cite{Fey63, Mak92, Sim01}
The diagonal elements of   $\left\{B_{kk'}\right\}$   describe local contributions to the bath response function from a  particular complex time slice $k$ along the adiabatic path,  and the off-diagonal  elements  describe  non-local contributions.
For the case of linear system-bath coupling, the diagonal matrix elements are given by
\begin{eqnarray}
	B_{kk}&&=\sum_{j=1}^f \frac{c_j^2}{M_j\omega_j^3
	\sinh(\beta\omega_j/2)} \nonumber\sin\left(\frac{\omega_j(t_{k+1}-t_{k})}{2}\right)\\
	&& \times\sin\left(\frac{\omega_j(t_{k+1}-t_{k}+i\beta)}{2}\right),
	\label{eq:bmat1}	
\end{eqnarray}	
and the off-diagonal matrix elements are given by
\begin{eqnarray}
	B_{kk'}&&=\sum_{j=1}^f \frac{c_j^2}{M_j\omega_j^3 
	\sinh(\beta\omega_j/2)} 	\sin\left(\frac{\omega_j(t_{k+1}-t_{k})}{2}\right) \nonumber \\
       &&\times\cos\left(\frac{\omega_j(t_{k+1}-t_{k'+1}+t_{k}-t_{k'}+i\beta)}{2}\right) \nonumber \\
	&&\times\sin\left(\frac{\omega_j(t_{k'+1}-t_{k'})}{2}\right).
	\label{eq:bmat2}
\end{eqnarray}
The complex times $t_k$ in Eqs.~\ref{eq:bmat1} and \ref{eq:bmat2} are provided in Table~\ref{tab:quapi_times}.

\begin{table}
\caption{Complex times $t_k$  used to calculate the $\left\{B_{kk'}\right\}$.
\label{tab:quapi_times}}
	\begin{ruledtabular}
	\begin{tabular}{cc}
\multicolumn{1}{c}{$k$}&\multicolumn{1}{c}{$t_k$}\\
\hline
$1$&$0$\\
$2, \ldots, \mathcal{N}+1$&$(k-1/2)\Delta t_\mathrm{c}$\\
$\mathcal{N}+2$&$t-i\beta\hbar/2$\\
$\mathcal{N}+3, \ldots, 2\mathcal{N}+2$&$(2\mathcal{N}+3/2-k)\Delta t_\mathrm{c}^*-i\beta\hbar$\\
$2\mathcal{N}+3$&$-i\beta\hbar$
	\end{tabular}
	\end{ruledtabular}
\end{table}

\subsection{Marcus Theory for ET in a Classical Solvent}

In the Marcus theory for ET,\cite{Mar85,Mar56,Mar60,Mar65} electronic transitions occur at solvent geometries for which the donor and acceptor electronic states are isoenergetic. In the limit of weak electronic coupling and classical solvent motions, the ET rate is thus
\begin{equation}
\label{eq:marcus_rate}
k_{\mathrm{\,MT}}=\frac{2\pi}{\hbar} |V_\mathrm{12}|^2 \left(\frac{\beta}{4\pi \lambda}\right)^{1/2} e^{-\beta \Delta G^{*}},
\end{equation}
where $V_\mathrm{12}$ is the electronic coupling matrix element,  
\begin{equation}
\label{eq:marcus_activation}
\Delta G^{*}=\frac{(\Delta G^0+\lambda)^2}{4 \lambda},
\end{equation}
 $\lambda$ is the solvent reorganization energy, and $-\Delta G^0$ is the thermodynamic driving force for the ET reaction. 
The rate expression in Eq.~\ref{eq:marcus_rate} exhibits three distinct regimes of behavior as the driving force is varied relative to $\lambda$. In the normal regime, where $-\Delta G^0 < \lambda$, the rate increases with increasing driving force. A turnover in this trend occurs in the activationless regime, for which $-\Delta G^0\approx\lambda$.  In the inverted regime, for which $-\Delta G^0 > \lambda$, the rate decreases with increasing driving force.

In the current study, we use implementations for Marcus theory, SCI
theory, and RPMD in which the solvent degrees of freedom are treated
classically; the role of nuclear quantum effects in diminishing the degree
of turnover for the ET rate in the inverted regime\cite{Mar85,Uls75} is not
considered here.

\section {Systems}
\label{sec:models}

ET dynamics is studied using 
both all-atom and system-bath representations for mixed-valence transition metal ions in water. These representations are described in this section. 

\subsection{Atomistic Representation for ET}
\label{sec:atomistic_model}

\label{sec:atomistic_potential}

\begin{figure}
\includegraphics{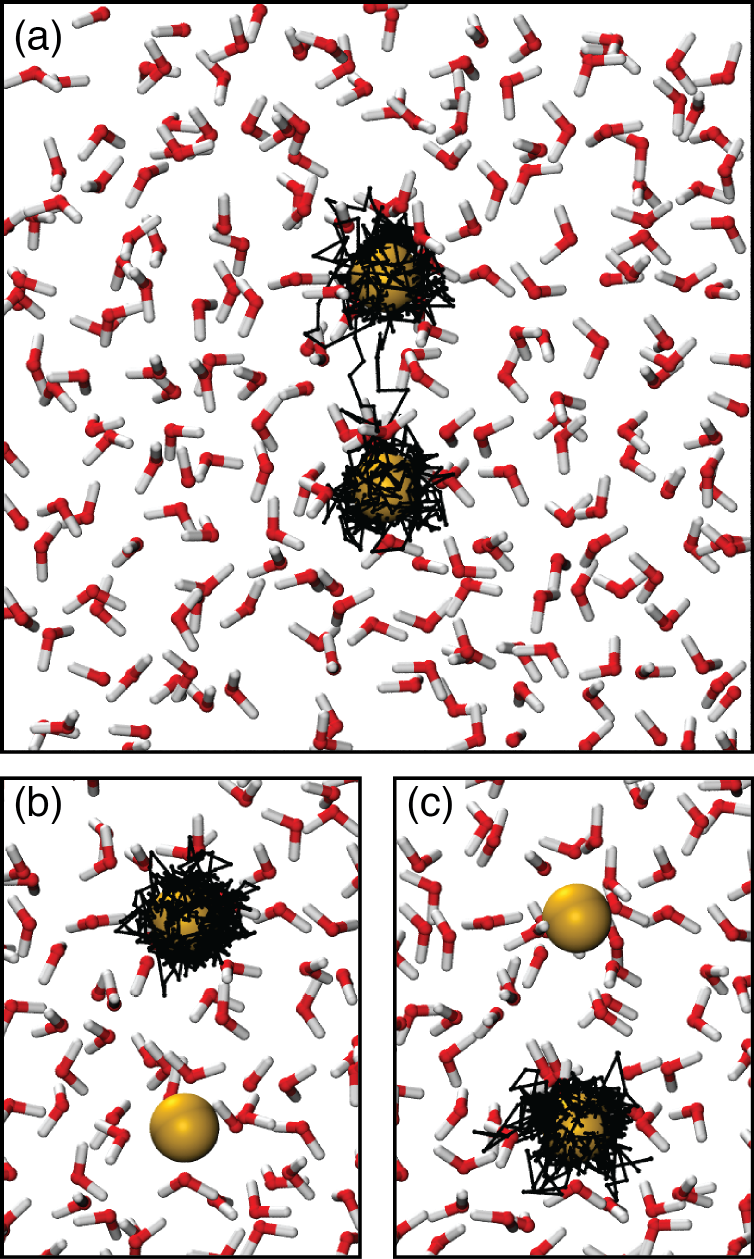}
\caption{
Snapshots of the atomistic representation for the ET reaction, with the donor and
acceptor metal ions shown in yellow, the electron ring polymer in black, and the
water molecules in red and white. Typical configurations of the symmetric ET system
are presented with the electron ring polymer (a) in transition between the redox
sites, (b) in the reactant basin, and (c) in the product basin.}
\label{fig:system}
\end{figure}

The atomistic representation for the 
ET reaction (Fig.~\ref{fig:system}) is described using the potential energy function\cite{Kuh88} 
\begin{align}
 U_\mathrm{ext}(\mathbf {q},\mathbf{Q})= U_{\mathrm{sol}}(&\mathbf{Q})+U_{\text{e-sol}}(\mathbf {q}^{},\mathbf{Q}) \nonumber \\
 &+U_{\text{e-M}}(\mathbf {q},\mathbf{Q})+U_{\text{M-sol}}(\mathbf{Q}),
\end{align}
where $\mathbf{q}$ is the electron position and $\mathbf{Q}$ is the set of $N$ classical solvent atom positions. 
Solvent-solvent interactions, $U_{\mathrm{sol}}(\mathbf{Q})$, are described using the  simple point charge (SPC) model \cite{Ber81} for explicit, rigid water molecules. The remaining interactions are described below, with the values of the parameters provided in~Table~\ref{tab:atomistic_params}.

\begin{table}
\caption{\label{tab:atomistic_params} Parameters for the atomistic representation of ET.} 
\begin{ruledtabular}
\begin{tabular}{cc}
\multicolumn{1}{c}{Parameter}&\multicolumn{1}{c}{Value}\\
\hline
$\mathbf{Q_\mathrm{A}}$ /\AA & $(0,0,-3.25)$  \\
$\mathbf{Q_\mathrm{D}}$ /\AA&  $(0,0,3.25)$\\
$r_{\mathrm{cut}}^\mathrm{H}$ /\AA& $1.0$\\
$r_{\mathrm{cut}}^\mathrm{M}$ /\AA&$1.1 $\\
$\mathrm{q_O}$ /$e$&$-0.84$\\
$\mathrm{q_H}$ /$e$&$0.42$\\
$\mathrm{q_M}$ /$e$&$3.0$\\
$\gamma_\mathrm{O}$ / (kcal/mol \AA$^9$)&$6392.7$\\
\end{tabular}
\end{ruledtabular}
\end{table}

The electron-water interactions are described using the pairwise pseudopotential \cite{Spr86}
\begin{equation}
U_{\text{e-sol}}\left(\mathbf {r}\right)=\sum_{k=1}^{N} U_{\text{e-sol}}^k \left( r_k\right),\nonumber 
\end{equation}
where $r_k=|\mathbf{q}-\mathbf{Q}_{k}|$.  For cases in which the atom index $k$ corresponds to a hydrogen atom, 
\begin{equation}
U^k_{\text{e-sol}}(r_k)=\begin{dcases} 
-\frac{\mathrm{q_H} e}{\pieps r_{\mathrm{cut}}^\mathrm{H}},&r_k \leq r_{\mathrm{cut}}^\mathrm{H} \\
-\frac{\mathrm{q_H} e}{\pieps r_k},& r_k > r_{\mathrm{cut}}^\mathrm{H},
\end{dcases}
\end{equation}
and when $k$ corresponds to an oxygen atom, 
\begin{equation}
\label {eq:atom_u_ewater2}
U^k_{\text{e-sol}}(r_k)=-\frac{\mathrm{q_O} e}{\pieps r_k}.
\end{equation}

Electron-ion interactions are described using
\begin{equation}
U_{\text{e-M}}(\mathbf {q})=U_{\text{e-D}}(|\mathbf{q}-\mathbf{Q}_ \mathrm{D}|)+U_{\text{e-A}}(|\mathbf{q}-\mathbf{Q}_ \mathrm{A}|),
\end{equation}
where $\mathbf{Q}_{\mathrm{D}}$ and $\mathbf{Q}_{\mathrm{A}}$ denote the respective positions of the donor and acceptor metal ions, which are held fixed at a separation of $6.5$~\AA.  These interactions are described using Shaw-type pairwise pseudopotentials.\cite{Sha68}  For the acceptor metal ion, 
\begin{equation}
\label {eq:atom_u_eredox}
U_{\text{e-A}}(r)=\begin{dcases} 
-\frac{\left( \mathrm{q}_{\mathrm{M}}+\epsilon \right) e }{\pieps r_{\mathrm{cut}}^\mathrm{M}},&  r  \leq r_{\mathrm{cut}}^\mathrm{M} \\
-\frac{\left( \mathrm{q}_{\mathrm{M}}+\epsilon\right) e }{\pieps r},& r > r_{\mathrm {cut}}^\mathrm{M},
\end{dcases}
\end{equation}
where $r=|\mathbf{q}-\mathbf{Q}_ \mathrm{A}|$, and for the donor metal ion, 
\begin{equation}
\label {eq:atom_u_eredox2}
U_{\text{e-D}}(r)=\begin{dcases} 
-\frac{ \mathrm{q}_{\mathrm{M}}e }{\pieps r_{\mathrm{cut}}^\mathrm{M}},& r \leq r_{\mathrm{cut}}^\mathrm{M}\\
-\frac{ \mathrm{q}_{\mathrm{M}}e }{\pieps r},& r > r_{\mathrm{cut}}^\mathrm{M},
\end{dcases}
\end{equation}
with $r=|\mathbf{q}-\mathbf{Q}_\mathrm{D}|$.
The asymmetry parameter, $\epsilon$, adjusts the thermodynamic driving force for the ET reaction while leaving the solvent reorganization energy unchanged. 
The values of $\epsilon$ considered in this study and the corresponding ET regimes are presented in Table~\ref{tab:atom_asym_values}.

\begin{table}
\caption{Values of the asymmetry parameter $\epsilon$ considered in the atomistic
representation and the corresponding thermodynamic driving force
regimes.
\label{tab:atom_asym_values}}
\begin{ruledtabular}
\begin{tabular}{ccc}
\multicolumn{1}{c}{Case}&\multicolumn{1}{c}{$\epsilon$ / $e$}&\multicolumn{1}{c}{ET Regime}\\
 \hline
I &  $0.0$ & Symmetric \\
II&$0.1$ &Normal \\
III&$0.2$ &Normal\\
IV &$0.3$&Activationless \\
V&$0.4$ &Inverted  \\
VI &$0.6$&Inverted  \\
VII &$0.7$&Inverted  \\
\end{tabular}
\end{ruledtabular}
\end{table}

The ion-water interactions are given by 
\begin{equation}
U_{\text{M-sol}}(\mathbf{Q})=\sum_{k=1}^{N}\left (U_{\text{D-sol}}^k(\mathbf{Q}_k)+U_{\text{A-sol}}^k(\mathbf{Q}_k) \right ).
\end{equation}
For cases in which atom index $k$ corresponds to a hydrogen atom,
\begin{equation}
U_{\text{D-sol}}^k(\mathbf{Q}_k)=\frac{\mathrm{q_H} \mathrm{q_M} }{\pieps |\mathbf{Q}_ \mathrm{D}-\mathbf{Q}_{k}|},
\label{EqUd1}
\end{equation}
and when  $k$ corresponds to an oxygen atom,
\begin{equation}
U_{\text{D-sol}}^k(\mathbf{Q}_k)=\frac{\gamma_\mathrm{O}}{|\mathbf{Q}_\mathrm{D}-\mathbf{Q}_{k}|^9}+ \frac{\mathrm{q}_{\mathrm{O} } \mathrm{q_M}}{\pieps|\mathbf{Q}_ \mathrm{D}-\mathbf{Q}_k |}.
\label{EqUd2}
\end{equation}
The potential energy functions associated with the acceptor ion,  $U_{\mathrm{A-sol}}^k(\mathbf{Q})$, are obtained by replacing $\mathbf{Q}_ \mathrm{D}$ with $\mathbf{Q}_ \mathrm{A}$ in Eqs.~\ref{EqUd1} and \ref{EqUd2}.
These ion-water potential energy functions include electrostatic interactions combined with short-range repulsive terms that reproduce the octahedral coordination structure of the solvated 
ions.\cite{Kuh88}

\subsection {System-Bath Representations for ET}
\label{sec:reduced_model}
The system-bath representation for the ET reaction is described  in the position basis using the potential energy function
\begin{equation}
\label{eq:sysbath}
U_{\mathrm{ext}}(q,s,\mathbf{Q})=U_{\mathrm{e-M}}\left(q\right)+U_{\mathrm{e-sol}}\left(q,s\right)+U_{\mathrm{B}}\left(s,\mathbf{Q}\right), 
\end{equation}
where the first two terms comprise the system potential, and $U_{\mathrm{B}}$ is the potential energy contribution due to the bath. The scalar coordinates $q$ and $s$ are the positions of the electron and the solvent mode, respectively.

The first term in the system potential energy function models the ion-electron interaction,
\begin{equation}
\label{eq:u_redox}
U_{\text{e-M}} (q)=\begin{dcases} 
a_\mathrm{L} q^2+b_\mathrm{L} q+c_\mathrm{L},&  r^{\mathrm{out}}_\mathrm{L} \leq q \leq r^{\mathrm{in}}_\mathrm{L}  \\
a_\mathrm{R} q^2+b_\mathrm{R}q+c_\mathrm{R},& r^{\mathrm{in}}_\mathrm{R} \leq q \leq r^{\mathrm{out}}_\mathrm{R} \\
- \frac{ (3 +\epsilon)}{|q-r_{\mathrm{A}}|} - \frac{3}{|q-r_{\mathrm{B}}|},& \textrm{otherwise}. 
\end{dcases}
\end{equation}
This one-dimensional (1D) potential energy function consists of two Coulombic wells capped by parabolic functions to remove the singularity; it is continuous, and its derivative is piecewise continuous over the full range  of $q$. The coefficients in Eq.~\ref{eq:u_redox} are provided in Appendix \ref{app:parabolic_caps}, and the values of $\epsilon$
considered for the system-bath representation are presented in Table~\ref{tab:model_asym_values}.

\begin{table}
\caption{Values of the asymmetry parameter $\epsilon$ considered in the
system-bath representation, the corresponding thermodynamic driving
force regimes, and the electronic coupling matrix element, $V_{12}$.\footnote{The coupling $|V_{12}|$ is given in units of $ \mathrm{a.u.} / 10^{7}$ for Model SB1 and $ \mathrm{a.u.} / 10^{5}$ for Model SB2; $\epsilon$ is in atomic units.}\label{tab:model_asym_values}}
\begin{ruledtabular}
\begin{tabular}{cccccc}
 \multirow{2}{*}{Case} &  \multicolumn{2}{c}{Model SB1}  &  \multicolumn{2}{c}{Model SB2}   & \multirow{2}{*}{ET Regime}\\
 &$\epsilon$&$|V_{12}|$& $\epsilon$& $|V_{12}|$ &\\
 \hline
 I & $0.0$ & $6.6860$&  $0.0$ & $2.0662$& Symmetric\\
II&$0.05$ &$6.4837$&  $-0.015$ & $2.0916$  &Normal\\
III&$0.10$ & $6.1300$& $-0.025$ & $2.1088$& Normal\\
IV &$0.20$ &$5.4840$&  $-0.050$ &  $2.1524$&Activationless\\
V&$0.30$ & $4.9120$&  $-0.075$ &  $2.1971$ &Inverted\\
VI &$0.40$ & $4.4040$& $-0.100$& $2.2427$ &Inverted \\
\end{tabular}
\end{ruledtabular}
\end{table}

The second term in the system potential energy function models the solvent and its interactions with the transferring electron, 
\begin{equation}
\label{eq:u_int}
U_{\text{e-sol}}({q},s) =  \mu s \tanh\left(\phi {q}\right) + \frac{1}{2}m_{\mathrm{s}}\omega_{\mathrm{s}}^2 s ^2.
\end{equation}
The first term on the RHS of this equation describes the coupling of the electronic dipole of the redox system to the solvent dipole, and $\omega_{\mathrm{s}}$ is the effective frequency of the solvent coordinate.  

The harmonic oscillator bath potential in Eq.~\ref{eq:u_redox} has the same form as in~Eq.~\ref{eq:quapi_bath_hamiltonian}, 
\begin{eqnarray}
\label{eq:u_sysbath}
U_{\mathrm{B}}(s,\mathbf{Q})&=&\sum_{j=1}^f  \left[ \frac{1}{2} M\omega_j^2 \left(Q_j-\frac{c_j s}{M \omega_j^2} \right)^2\right].
\end{eqnarray}
The bath exhibits Ohmic spectral density with cutoff frequency $\omega_{\mathrm{c}}$,
\begin{equation}
J(\omega)=\eta \omega e^{-\omega/\omega_{\mathrm{c}}},
\end{equation}
where the dimensionless parameter $\eta$ determines the strength of coupling between the system and the bath modes.\cite{Cal83} The continuous spectral density is discretized into $f$ oscillators with frequencies \cite{Cra05}
\begin{equation}
\omega_j=-\omega_{\mathrm{c}}\log\left(\frac{j-0.5}{f}\right)
\end{equation}
and coupling constants 
\begin{equation}
c_j=\omega_j\left(\frac{2\eta M \omega_{\mathrm{c}}}{f\pi}\right)^{1/2},
\end{equation}
where $j=1\ldots f$.

In the current paper, we use two sets of parameters for the system-bath representation. Model SB1 is constructed to reproduce the energy-scales of  the atomistic representation, and Model SB2 uses parameters that are numerically less demanding for the QUAPI calculations. The parameters for the models are given in Table \ref{tab:reduced_params}. 

\begin{table}
\caption{Parameters for the system-bath representation of ET.\footnote{Parameters  given in atomic units, unless otherwise specified.}\label{tab:reduced_params}}
\begin{ruledtabular}
\begin{tabular}{ccccc}
 Parameter & Model SB1 & Model SB2 \\
\hline
$r_{\mathrm{A}}$/\AA & $3.25$  & $2.72435$ \\ 
$r_{\mathrm{B}}$/\AA &  $-3.25$ &  $-2.72435$ \\
 $\mu$& $0.0230725$ & $0.0114265$ \\
 $f$ & \multicolumn{2}{c}{ $12$ }    \\
$\omega_{\mathrm{s}}$ &  \multicolumn{2}{c}{ $0.00228$} \\
$\omega_{\mathrm{c}}$ &  \multicolumn{2}{c}{ $0.00228$}  \\
$M$  &   \multicolumn{2}{c}{ $1836.0$ }\\
$m_{\mathrm{s}}$ & \multicolumn{2}{c}{ $1836.0$ } \\
$\eta/ M \omega_{\mathrm{c}} $ &  \multicolumn{2}{c}{ $1.0$ }\\
 $\phi/ r_{\mathrm{A}}$ & \multicolumn{2}{c}{ $3.0 $}  \\
\end{tabular}
\end{ruledtabular}
\end{table}

As indicated previously, the QUAPI method is implemented using a discrete representation for the diabatic states of the redox system (Eq.~\ref{eq:tls_hamiltonian}). 
The system representation in the position basis described in Eq.~\ref{eq:sysbath} is therefore transformed to the electronic diabatic basis for the QUAPI calculations. The resulting diagonal matrix elements for the system potential energy are
\begin{equation}
\label{eq:v11}
V_{11}(s)=a_1s^2+b_1s+c_1   
\end{equation}
and 
\begin{equation}
\label{eq:v22}
V_{22}(s)=a_2s^2+b_2s+c_2,
\end{equation} 
and the constant off-diagonal elements $V_{12}$ are reported in Table~\ref{tab:model_asym_values}. The details of this transformation and the values of the coefficients in Eqs.~\ref{eq:v11} and \ref{eq:v22} are given in Appendix \ref{app:diabatic_localization}.

\section{Calculation Details}
\label{sec:calc_details}
\subsection{Atomistic Representation}
\label{cdAtomistic}

The atomistic system includes 430 SPC water molecules in a cubic simulation cell with periodic boundary conditions. The side-length of the cell is $L=23.46$~\AA. All calculations are performed at a temperature of $T=300$ K, and all pairwise interactions are truncated at  a distance of ${r_{\mathrm{cut}}=L/2}$. Long-range electrostatics  are treated by the force-shifting algorithm,\cite{Bro85} where the Coulombic portion of each potential is multiplied by a damping function $S(r)$, such that both the potential and its derivative smoothly
vanish at ${r=r_{\mathrm{cut}}}$.  Specifically,
\begin{equation}
S(r)=\begin{dcases} 
1-\frac{2r}{r_{\mathrm{cut}}}+\frac{r^2}{{r_{\mathrm{cut}}}^2},& r \leq r_{\mathrm{cut}} \\
0, & r > r_{\mathrm{cut}}
\end{dcases}.
\end{equation}
Force-shifting reduces the unphysical structuring of water near the the cutoff radius,\cite{Bro85} and it is found to have little effect on the solvent environment of the redox system.

\subsubsection{RPMD}

The atomistic RPMD simulations are implemented in the DL\_POLY molecular dynamics package. \cite{dlpoly2} In all simulations, the RPMD equations of motion 
are evolved using the velocity Verlet algorithm, \cite{Ver67} and the constraints in the rigid-body water model are implemented using the RATTLE algorithm.\cite{And83}  The electron is quantized with ${n=1024}$ ring-polymer beads. 
As in previous RPMD simulations, each timestep for the electron ring polymer involves separate coordinate updates due to forces arising from the physical potential and due to exact evolution of the purely harmonic portion of the ring-polymer potential. The resulting integration algorithm is time-reversible and symplectic.

Several collective variables are used to characterize the ET reaction in the atomistic representation. The position of the electron is described by a ring-polymer progress variable, or ``bead-count" coordinate,  defined as
\begin{equation}
\label{eq:fb}
f_\mathrm{b}(\mathbf{q}^{(1)},\ldots,\mathbf{q}^{(n)})=\frac{1}{n}\sum_{\alpha=1}^n \frac{1}{2}\left(\tanh \left(b q_z^{(\alpha)} \right)+1\right),
\end{equation}
where $b=1.25$ \AA$^{-1}$, and the metal ions are symmetrically positioned on the $z$-axis. We also consider the solvent  collective variable
\begin{equation}
\Delta U(\mathbf{Q})=-\frac{e}{\pieps}\sum_{k=1}^{N}\left (\frac{\mathrm{q_k} }{|\mathbf{Q}_ \mathrm{D}-\mathbf{Q}_{k}|}-\frac{\mathrm{q_k}}{|\mathbf{Q}_ \mathrm{A}-\mathbf{Q}_{k}|}\right),
\label{SolventVariable}
\end{equation}
where $q_k\in\left\{\mathrm{q_H},\mathrm{q_O}\right\}$ is the charge on solvent atom $k$.  
This solvent collective variable, which is familiar from earlier simulation studies of Marcus theory,\cite{Kin90, Kuh88} describes the energy difference between the electonic diabatic states in the tight-binding approximation.

The RPMD rate in Eq.~\ref{eq:rpmd_rate} is calculated from the product of the TST rate  and the transmission coefficient.
The TST rate described in Eq.~\ref{eq:rpmd_tst} is obtained from $F(f_{\mathrm{b}})$, the FE profile in the bead-count coordinate. 
This FE profile is calculated using umbrella sampling and the weighted histogram analysis method (WHAM), as described below.\cite{Kum92,Kum94}

For each value of the asymmetry parameter $\epsilon$, the following umbrella sampling protocol is used.
The region $f_\mathrm{b}=0.06-0.94$ is sampled with $22$ trajectories that are harmonically restrained to uniformly spaced values of $f_\mathrm{b}$ using a  restraint force constant of $1.195\times10^{4} $ kcal/mol. Likewise, the regions  $f_\mathrm{b}=0.945-1.0$ and $f_\mathrm{b}=0.0-0.055$ are each sampled with 11 uniformly spaced windows using a higher force constant of $1.195\times10^{5}$~kcal/mol. The regions of $f_\mathrm{b}=0.986-0.991$ and $f_\mathrm{b}=0.009-0.015$ are each sampled with 5 uniformly spaced windows using a force constant of $1.195\times10^{5}$ kcal/mol.  The equilibrium sampling trajectories are performed using path-integral molecular dynamics (PIMD) with a Parrinello-Rahman mass of $364.6$~a.u., which allows for a timestep of $0.025$~fs; this choice of mass does not affect the calculated FE profile or any other equilibrium ensemble average.\cite{Par84,DeR84} Each sampling trajectory is run for at least $50$~ps, and thermostatting is performed during the trajectory calculations by resampling the particle velocities from the Maxwell-Boltzmann (MB) distribution every $1.25$~ps.

The transmission coefficient in Eq.~\ref{eq:kappa_final} is calculated 
using RPMD trajectories that are released from the dividing surface at $f_{\mathrm{b}}^{\ddagger}$.
For each value of $\epsilon$,  the dividing surface is chosen to coincide with the maximum along the FE profile, $F(f_{\mathrm{b}})$. The positions of the dividing surfaces are set to $f_{\mathrm{b}}^{\ddagger}=(0.5,~0.7,~0.8,~0.96,~0.98,~0.98,~0.98)$ for the different $\epsilon$-cases (I,~II,~\dots,~VI) in Table~\ref{tab:atom_asym_values}. Between $400$ and $1200$ trajectories are released for each value of $\epsilon$. Each RPMD trajectory is evolved for $40$~fs with a timestep of $5\times10^{-5}$~fs and with the initial velocities sampled from the MB distribution. 
Initial configurations for the released RPMD trajectories are selected every $100$~fs from eight long, independent PIMD sampling trajectories that are constrained to the dividing surface $\xi^{\ddagger}$.  These sampling trajectories are thermostatted by resampling the velocities every $200$~fs, and the
constraint to the dividing surface is enforced using the RATTLE algorithm.

Two-dimensional (2D) FE surfaces in the ring-polymer centroid coordinate and the solvent coordinate, $F(\bar{z},\Delta U)$, are used for the analysis of the ET reaction mechanism.  For a given value of $\epsilon$, the 2D FE surface is constructed using PIMD sampling trajectories that are harmonically restrained in both $\bar{z}$ and $\Delta U$ coordinates. The $\bar{z}$ coordinate is sampled using $43$ uniformly spaced windows in the region of $-3.575$  \AA~to $+3.575$~\AA~with a harmonic restraint force constant of $169.7$ kcal/mol \AA$^{-2}$.  To ensure adequate sampling of ring-polymer configurations spanning both metal ions, we use four additional sampling trajectories that are harmonically restrained to $\bar{z}=\pm2.7625$~\AA~and $\bar{z}=\pm2.925$~\AA~  with a force constant of $452.5$ kcal/mol  \AA$^{-2}$. The solvent coordinate is sampled with $15$ uniformly spaced windows in the range $-130~\mathrm{to}~+150$~kcal/mol using a harmonic restraint force constant of $0.023$~(kcal/mol)$^{-1}$.  Each sampling trajectory is run for at least $50$~ps, with velocities resampled from the MB distribution every $500$ fs.
We note that $f_{\mathrm{b}}$ is a good progress variable for ET throughout the entire regime of the thermodynamic driving forces, whereas the ring-polymer centroid is not.  In the ET inverted regime, the centroid does not fully distinguish between ring-polymer configurations in the reactant and product basins; no such difficulty is experienced in the calculations reported here. 

\subsubsection{Marcus Theory}
\label{cdMarcusTheory}
Marcus theory rates are calculated using Eqs.~\ref{eq:marcus_rate}~and~\ref{eq:marcus_activation}. The driving force, $-\Delta  G^0$, is obtained from $F(\Delta U)$  as the difference between the free energies of the reactant and product minima; these values are reported in Table~\ref{tab:atomistic_rates}. 
To the extent that the tight-binding approximation holds, the reorganization energy, $\lambda$, is identical for all $\epsilon$, and we confirm that this is very nearly the case in our calculations. For the case of symmetric ET ($\epsilon=0$), the reorganization energy is calculated using $\lambda=4F(\Delta U)|_{\Delta U=0}$ and is found to be $69.7\pm 0.7$~kcal/mol.

The coupling matrix element in Eq.~\ref{eq:marcus_rate}, $|V_{\mathrm{12}}|$, is calculated as ${2|V_{\mathrm{12}}|=E_1-E_0}$, where $E_0$ and $E_1$  are the two lowest eigenenergies of the electron in the potential of the isolated metal ions with ${\epsilon=0}$. These eigenenergies are obtained with an iterative, block Lanczos scheme, \cite{Web91} performed on a uniform grid of ${64\times64\times64}$ points spanning the 
cubic simulation cell. The iterative Lanczos calculation employs $200$ Krylov vectors and an exponential transform parameter of ${\beta_{\mathrm{L}}=0.1}$. The block Lanczos refinement uses ten blocks of five Krylov vectors. This yields a value for the tunnel splitting of $|{V_{\mathrm{12}}|=0.0403}$~kcal/mol~($6.43\times10^{-5}$~a.u.), which is consistent with previous calculations.\cite{Kuh88}
This value for the tunnel splitting was assumed to be insensitive to presence of solvent, 
as has been previously demonstrated,\cite{Mar91}, and independent of the value of the asymmetry parameter $\epsilon$. 
The validity of this latter assumption is confirmed for the system-bath models (see Table \ref{tab:model_asym_values}).

\subsection{System-Bath Representation}
As in the atomistic representation, the calculations in the system-bath representation are performed at ${T=300}$~K. The harmonic bath is discretized using ${f=12}$  modes.

\subsubsection{RPMD}
RPMD rates for the system-bath models are also calculated 
with the electron quantized using ${n=1024}$ ring-polymer beads. For each value of $\epsilon$, the FE profile, $F(f_{\mathrm{b}})$, is obtained from umbrella sampling  along the $f_{\mathrm{b}}$ coordinate. For both system-bath models, SB1 and SB2, $F(f_{\mathrm{b}})$ is sampled with two sets of harmonically restrained PIMD trajectories. The region of $f_{\mathrm{b}}=0.06-0.94$ is sampled with $45$ trajectories that are harmonically restrained to uniformly spaced values of $f_{\mathrm{b}}$ using a force constant of $20$ a.u. The regions of $f_{\mathrm{b}}=0.0-0.05$ and $f_{\mathrm{b}}=0.095-1.00$ are each sampled with $51$ uniformly spaced windows using a harmonic restraint force constant of $3000$~a.u.  All sampling trajectories are performed using PIMD  with the masses of the classical particles set to $m_{\mathrm{s}}=M=0.01$~a.u; as before, the altered masses in the PIMD sampling trajectories allow for larger timesteps while having no effect on calculated ensemble averages. Each sampling trajectory is run for at least $12.09$~ps, the PIMD timestep is $2.42\times10^{-4}$~fs, and thermostatting is performed by resampling velocities  from the MB distribution every $2.42$~fs.  The FE profiles are constructed from the sampling trajectories using WHAM.

For each value of $\epsilon$, 
the transmission coefficient  in Model SB1 is calculated from $2400$ RPMD trajectories released from the dividing surface and evolved for $121$~fs with the timestep of ${1.21\times10^{-4}}$~fs. The position of the dividing surface is ${f_{\mathrm{b}}^{\ddagger}=(0.5,~0.385,~0.2345,~0.014,~0.014,~0.014)}$ for the $\epsilon$-cases (I,~II,~\dots,~VI).  In Model SB2, $1600$ RPMD trajectories are released at each value of $\epsilon$; each trajectory is evolved for $121$~fs using a timestep of ${2.42\times10^{-4}}$~fs; and the dividing surface is located at ${f_{\mathrm{b}}^{\ddagger}=(0.5,~0.65,~0.75,~0.986,~0.986,~0.986)}$ for $\epsilon$-cases (I,~II,~\dots,~VI). 
  Initial configurations for the released RPMD trajectories are sampled every $14.5$~fs from eight long, independent 
PIMD sampling trajectories that are constrained to the dividing surface. The velocities of the PIMD sampling trajectories are resampled every  $48.4$~fs from the MB distribution.
 The dividing surface constraint is implemented using the RATTLE algorithm. 
 
We note that RPMD results can be affected by coupling of fictitious internal ring-polymer modes to physical frequencies in the system.\cite{Hab08}  We thus performed test calculations of the ET rate in these and similar systems using partially adiabatic centroid molecular dynamics (PACMD).\cite{Hab08, Hon06} The PACMD calculations revealed no significant changes from the RPMD results, confirming that this issue does not impact our conclusions.

\subsubsection{Marcus Theory}

For the calculation of Marcus theory rates, the reorganization energy and the thermodynamic driving force for each value of epsilon are obtained analytically from the diabatic states for the donor and acceptor, $V_{11}(s)$ and $V_{22}(s)$. 
For Model SB1, we obtain a solvent reorganization energy of ${\lambda=68.9}$~kcal/mol, and for Model SB2, we obtain ${\lambda=17.0}$~kcal/mol. The values of $|V_{12}|$ for both system-bath models is given in Table~\ref{tab:model_asym_values}.

\subsubsection{Semiclassical Instanton Theory}

For the SB models, contributions from the linearly-coupled harmonic bath can be factorized and cancelled from the RHS of Eq.~\ref{eq:sci_rate}, yielding expressions that depend only on the electron ring-polymer coordinates and the single classical solvent coordinate, $s$.  
Calculation of $k_{\mathrm{SCI}}$ 
then consists of 
\emph{(i)} determination of saddle-point configurations for the classical action, $A\left(\{q^{(\alpha)}\};s\right)$, on a numerical grid in the solvent coordinate $s$, 
\emph{(ii)} 
evaluation of the steepest-descent approximation for $\mathcal{I}_n(s)$ 
at each point on the solvent grid,
and
\emph{(iii)} integration over the solvent coordinate in Eq.~\ref{eq:sci_rate_sd} via numerical quadrature.  The reactant partition function, $Q_r$, was similarly obtained by evaluating $I_n(s)$ via steepest-decent expansion around the minimum-action configuration in the reactant basin. 
All calculations were performed using $n=2048$ beads for the electron ring polymer.

For Model SB1, the grid in the solvent coordinate $s$ consists of $200$ uniformly spaced points in the range of $-4$ to $4$ a.u.;  for Model SB2, this grid consists of 150 uniformly spaced points in the range $-3$ to $3$ a.u.  
At each value of $s$,
the saddle-point configuration on the surface $A\left(\{q^{(\alpha)}\};s\right)$ corresponds to the maximum  
along the path of minimum action that connects 
the reactant and product basins. 
This  path of minimum action is obtained using the string method,\cite{E07} with the path discretized into $L=1000$ equidistant slices and with minimization performed using Euler integration and a timestep of $2.4\times10^{-3}$~fs.  Initial convergence of the path is achieved when this minimization results in a change of less than $5.3\times10^{-8}$~\AA\ in each degree of freedom.
The path is then iteratively refined in the vicinity of the saddle point: a 20-slice sub-section of the path about the saddle point is extracted, the number of slices used to describe the path is doubled, and the sub-section of the path is re-minimized with its endpoints fixed.
Iterative refinement of the path is complete when the slice of maximum action (i.e., the saddle point configuration) satisfies Eq.~\ref{eq:sci_stationary} to within 
$10^{-5}$ a.u.

\subsubsection{QUAPI}

The QUAPI calculation for Model SB2
requires
construction of the short-time system propagator followed by two independent Monte Carlo (MC) simulations to evaluate the flux-flux correlation function in Eq.~\ref{eq:cff_quapi}. 

	The  complex-time propagator in Eq.~\ref{eq:qm_prop} is calculated using  eigenvalues and eigenfunctions obtained from a 2D discrete variable representation (DVR) grid calculation\cite{Col92} in the solvent coordinate, $s$, and the electronic state variable, $\sigma$. The DVR Hamiltonian is diagonalized on a grid of $40$ uniformly spaced points over a range of $-4$~to~$+4$~a.u. in $s$ and $\sigma=1,2$. The number of eigenvalues and eigenvectors used in these calculations ($M_0$ in Eq.~\ref{eq:qm_prop}) ranges from $30$ to $50$ for the values of $\epsilon$ considered in this study. 

The flux-flux correlation function in Eq.~\ref{eq:cff_quapi} is obtained from standard path-integral Monte Carlo (PIMC) sampling performed on the 2D DVR grid. 
In a first PIMC simulation,  the correlation function is obtained using
\begin{eqnarray}
	C_{\mathrm{FF}}(t)&=&D_\rho\left\langle\;\text{sgn}\{\text{Re}\left[
	I_1(\mathbf{s},\boldsymbol{\sigma};t_\mathrm{c})-I_2(\mathbf{s},\boldsymbol{\sigma};t_\mathrm{c}) \right.\right. \nonumber \\
	&+&\left.\left.  I_3(\mathbf{s},\boldsymbol{\sigma};t_\mathrm{c})-I_4(\mathbf{s},\boldsymbol{\sigma};t_\mathrm{c})\right]\}\;
	\right\rangle_{\rho(\mathbf{s},\boldsymbol{\sigma};t_\mathrm{c})},
\end{eqnarray}
where importance sampling is performed using the distribution 
\begin{eqnarray}
\rho(s,\sigma;t_\mathrm{c})&=&\text{Abs}\left\{\text{Re}
\left[I_1\mathbf{s}(\boldsymbol{\sigma};t_\mathrm{c})-I_2(\mathbf{s},\boldsymbol{\sigma};t_\mathrm{c}) \right.\right. \nonumber \\
&+&\left. \left. I_3(\mathbf{s},\boldsymbol{\sigma};t_\mathrm{c})-I_4(\mathbf{s},\boldsymbol{\sigma};t_\mathrm{c})
\right]\right\}, 
\end{eqnarray}
and the function  $I_i(\mathbf{s},\boldsymbol{\sigma};t_\mathrm{c})$ is defined in Eq.~\ref{integrandQuapi}. Convergence is achieved with $10^{8}$ MC steps. The normalization constant, $D_\rho$,
is obtained from a second, independent PIMC simulation, using
 \begin{equation}
	D_\rho=D_\Lambda \left\langle\;\frac{\rho(\mathbf{s},\boldsymbol{\sigma};t_\mathrm{c})}
	{\Lambda(\mathbf{s},\boldsymbol{\sigma};t_\mathrm{c})}\;
	\right\rangle_{\Lambda(\mathbf{s},\boldsymbol{\sigma};t_\mathrm{c})}.
\end{equation}
Here, importance sampling is performed on the distribution
\begin{eqnarray}
\Lambda(\mathbf{s},&&\boldsymbol{\sigma};t_\mathrm{c})=\prod_{k=\mathcal{N}+3}^{2\mathcal{N}+2}\left| \bra{\sigma_k,s_k}e^{iH_{\mathrm{S}}\Delta t_\mathrm{c}^*/\hbar}\ket{\sigma_{k-1},s_{k-1}} \right| \nonumber\\
&&\times\prod_{k=2}^{\mathcal{N}+1} \left| \bra{\sigma_k,s_k}e^{-iH_{\mathrm{S}}\Delta t_\mathrm{c}/ \hbar}\ket{\sigma_{k-1},s_{k-1}} \right|,
\end{eqnarray}
where ${\sigma_1=2}$, ${\sigma_{\mathcal{N}+1}=2}$, ${\sigma_{\mathcal{N}+2}=1}$, and ${\sigma_{2\mathcal{N}+2}=1}$. Convergence is achieved with  $10^{6}$ MC steps, and the normalization constant $D_ \Lambda $ is obtained by direct matrix multiplication.  A maximum of $\mathcal{N}=4$ path beads are required to converge the flux-flux correlation function over a timescale of $25$~fs; no significant changes are observed between calculations performed using $\mathcal{N}=4$ and $\mathcal{N}=8$. 

The reactant partition function is obtained from a single PIMC calculation using the expression
\begin{equation}
	Q_{\mathrm{R}} = \mathrm{Tr}[e^{-\beta H} \mathcal{P}_{\mathrm{1}}],
	\label{eq:qm_qr_defn}
\end{equation}
where $\mathcal{P}_\mathrm{1}=\ket{1}\bra{1}$ is the projection operator for the reactant electronic state.

The QUAPI calculations for case IV are performed using a larger value for the  coupling between the solvent coordinate and the bath modes, $\eta/M\omega_c=30$.  This change 
leads to lower-amplitude oscillations in the flux-flux correlation function and improved numerical convergence of the ET rate calculation.
Other features of the flux-flux correlation function, including the timescales for the real-time oscillations and the decorrelation time, are unchanged.
The invariance of these features 
suggests that the parameters used in the current study correspond to the regime in which the ET reaction rate is independent of the solvent-bath coupling.\cite{Top96,Cli88} RPMD rate calculations performed using different values for $\eta/M\omega_c$ also support this conclusion.

%
\section{Results}
\label{sec:results}

\subsection{Atomistic Simulations}

\begin{figure}
\includegraphics{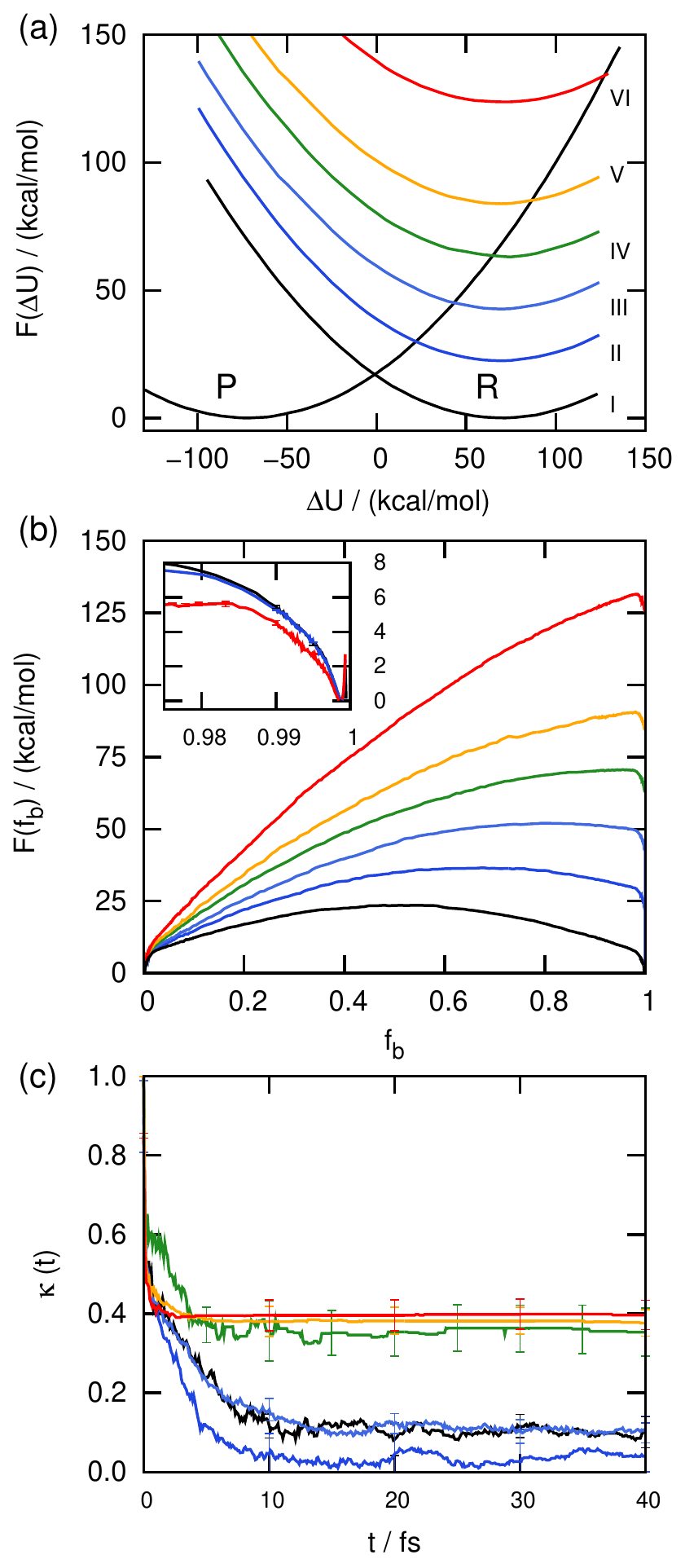}
\caption{\textrm{(a)} FE profiles, $F (\Delta U)$,  for the reactant (colored, at
right) and product (left) diabatic electronic states as a function of the solvent
collective variable in the atomistic representation. The various cases of the
thermodynamic driving force for the ET reaction are labeled; see Tables
\ref{tab:atom_asym_values} and \ref{tab:atomistic_rates} for details. For each case,
the FE profiles are vertically shifted to align the minima of the product basin.
\textrm{(b)} The corresponding FE profiles as a function of the bead-count
coordinate, $F (f_{\mathrm{b}})$.  In the main panel, the profiles are vertically
shifted to align the product basin; in the inset, the profiles are vertically
shifted to align the reactant basin.
\textrm{(c)} The corresponding RPMD transmission coefficients for the ET reaction,
$\kappa (t)$. 
In panels (b) and (c), the curves retain the same color scheme introduced in 
panel (a).
}
\label{fig:atom_fes}
\end{figure}

The atomistic representation for ET (Fig.~\ref{fig:system}) is investigated using direct RPMD simulations and the Marcus rate theory.
For each case of the thermodynamic driving force, Fig.~\ref{fig:atom_fes}(a) presents FE profiles for the reactant and product diabatic electronic states as a function of the solvent collective variable, $\Delta U(\mathbf{Q})$ (Eq.~\ref{SolventVariable}).  The FE profiles are obtained by reducing the corresponding 2D surfaces, $F(f_{\mathrm{b}},\Delta U)$, where the reactant and product diabats are associated with ring-polymer configurations for which ${f_{\mathrm{b}}>0.995}$ and ${f_{\mathrm{b}}<0.005}$, respectively.  
The results in Fig.~\ref{fig:atom_fes}(a) are graphically identical to those obtained using the tight-binding approximation, and the FE profiles exhibit the anticipated parabolic form, although no assumptions regarding the linear response of the solvent have been made. \cite{Kuh88,Kin90}  These data, in combination with the calculated tunnel splitting for the transferring electron, are used to calculate the Marcus rates in Table \ref{tab:atomistic_rates}. 

\begin{table}
\caption{ET reaction rates for the atomistic representation, obtained using RPMD and Marcus theory.\footnote{ET rates are given in $s^{-1}$, and the $-\Delta G^0$ are given in kcal/mol. The numbers in parentheses denote the statistical uncertainty in the last reported digit.} \label{tab:atomistic_rates}}
\begin{ruledtabular}
\begin{tabular}{ccdd}
Case &  $-\Delta G^o$& \multicolumn{1}{c}{$\log k_{\mathrm{MT}}$} & \multicolumn{1}{c}{$\log k_{\mathrm{RPMD}} $} \\
\hline
I   & $0.0$ &  -2.2(1) &      -1.7(2)\\
II & $22(1)$ & 4.6(3)     & 4.6(4)  \\
III  & $43(1)$ & 8.7(2)   &  8.4(2)  \\
IV &  $63(1)$ &10.40(4) & 10.17(9) \\
V  & $ 84(1)$  &  10.0(1)&  11.21(4) \\
VI  &  $124(3)$ & 2.9(8)  &11.48(8) \\
VII  & $138(2)$ & -1.8(9)  &11.80(7)  \\
\end{tabular}
\end{ruledtabular}
\end{table}

 Fig.~\ref{fig:atom_fes}(b) presents the corresponding FE profiles as a function of the bead-count coordinate, $f_\mathrm{b}$ 
  (Eq.~\ref{eq:fb}). 
These profiles are used in the statistical component of the RPMD rate calculation (Eqs.~\ref{eq:rpmd_rate} and \ref{eq:rpmd_tst}).  As is seen from the inset, all of the profiles behave similarly in the vicinity of $f_\mathrm{b}\approx 1$.  The steep rise in the FE profile between $0.980$ and $0.999$ is associated with the formation of ``kink-pair" configurations, in which the ring polymer spans both redox sites;\cite{Cha81,Chandler_LesHouches,Cep95}  a typical kink-pair configuration is illustrated in Fig.~\ref{fig:system}(a).

The dynamical component of the RPMD rate calculation (Eq.~\ref{eq:kappa_final}) is obtained from the long-time plateau\cite{Cha78} 
of the RPMD transmission coefficient shown in Fig.~\ref{fig:atom_fes}(c).  Each transmission coefficient is calculated with respect to a dividing surface at a fixed value of $f_\mathrm{b}$, as is described in Sec.~\ref{cdAtomistic}. Plateau values in the range of 0.1-0.4 indicate modest recrossing of the RPMD trajectories through these surfaces.  For cases in which the thermodynamic driving force corresponds to 
ET in the normal and acitivationless regimes, Fig.~\ref{fig:atom_fes}(c) illustrates that the RPMD trajectories commit to the reactant or product basins within 10-20 fs, 
the timescale for local solvent 
 motion between librational rebounds.  
 At thermodynamic driving forces corresponding to the inverted regime, the transmission coefficient plateaus on faster timescales than those involving the rigid solvent molecules.

\begin{figure}
\includegraphics{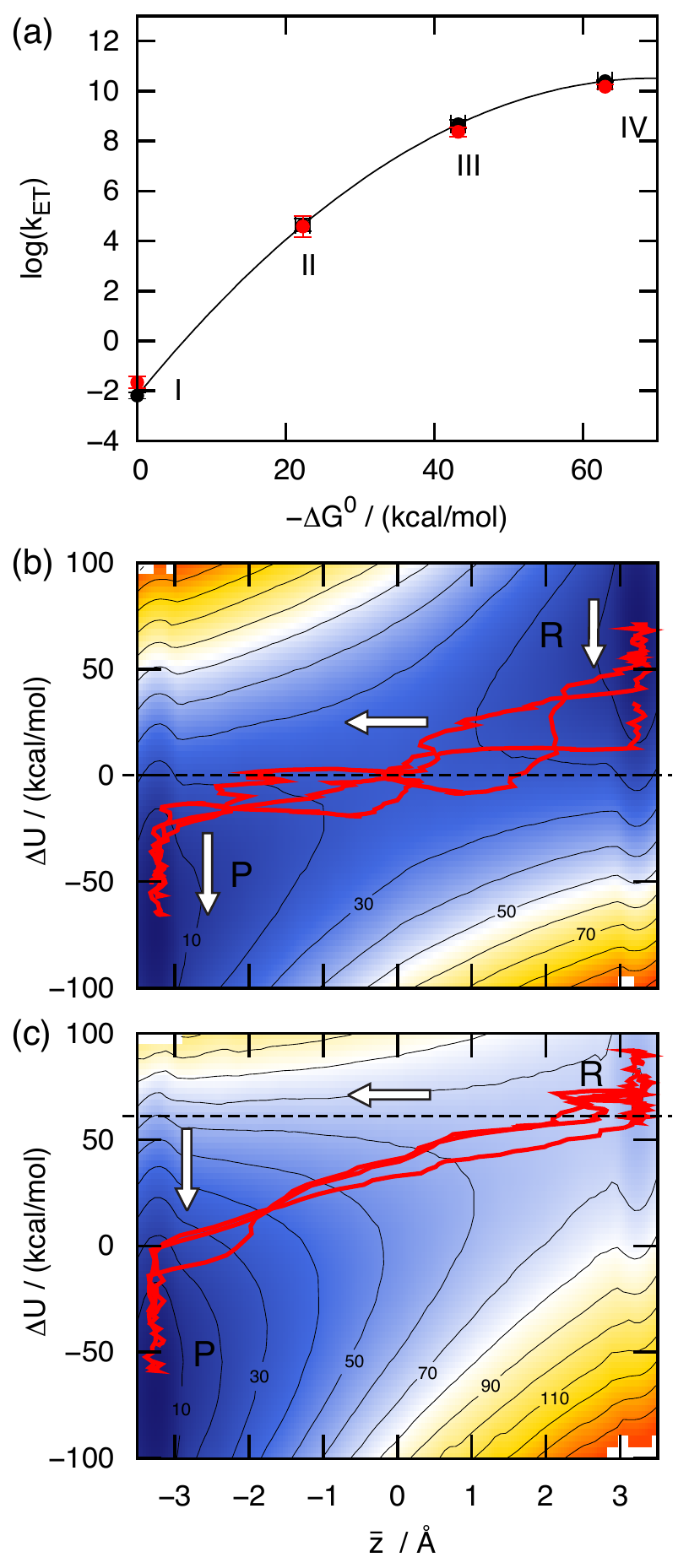}
\caption{(a)
ET reaction rates for the atomistic representation in the normal and activationless
regimes, computed using RPMD (red) and Marcus theory (black).  The various cases
for the thermodynamic driving force are labeled.
(b)  Representative trajectories (red) from the ensemble of reactive RPMD
trajectories for symmetric ET (Case I).  The trajectories are plotted as a function
of the ring-polymer centroid, $\bar{z}$, and the solvent collective variable,
$\Delta U$. The FE profile in these collective variables is also presented, with
contour lines indicating FE increments of 10 kcal/mol.
(c) Representative RPMD trajectories for activationless ET (Case IV) and the
corresponding FE profile.
The white arrows in panels (b) and (c) indicate the solvent reorganization mechanism
for ET that is anticipated in the Marcus rate theory, and the dashed lines indicate
values of $\Delta U$ at which the reactant and product diabats cross in
Fig.~\ref{fig:atom_fes}(a).
\label{fig:atom_normal}}
\end{figure}

Fig.~\ref{fig:atom_normal}(a) presents a direct comparison of the RPMD and Marcus theory rates throughout the normal and activationless regime for ET in the atomistic representation.  The RPMD rates, which are also reported in Table \ref{tab:atomistic_rates}, quantitatively agree with the Marcus theory results over 12 orders of magnitude in the ET reaction rate.
Unlike the Marcus rates, which are based on a TST description for the reaction, the calculated RPMD rates are independent of any \emph{a priori} assumptions about the ET reaction mechanism.

Figs.~\ref{fig:atom_normal}(b) and (c) illustrate the ET reaction mechanism that is predicted from the RPMD simulations. 
Representative RPMD trajectories are projected onto the $(\bar{z}, \Delta U)$ plane, where $\bar{z}$ is the component of the ring-polymer centroid that lies along the axis of the metal ions in the system;
  also shown are FE profiles for the system in these collective variables.
  For symmetric ET (Case I),  Fig.~\ref{fig:atom_normal}(b) reveals that the RPMD trajectories involve three distinct steps that will be familiar from the Marcus rate theory: \emph{(i)} solvent fluctuation to a configuration for which the reactant and product diabats are nearly degenerate (indicated by the dashed line), \emph{(ii)} formation of a kink-pair in the ring-polymer configuration and rapid transfer of the electron from one redox site to the other, and \emph{(iii)} relaxation of the solvent coordinate in the product basin following the ET event.
  For ET approaching the activationless regime (Case IV), Fig.~\ref{fig:atom_normal}(c) shows the latter two steps in the mechanism remain, but only a small initial solvent fluctuation is needed to reach solvent configurations for which the electronic diabats are degenerate.
  
To understand the connection between RPMD and the Marcus theory rate expression, we note that Eq.~\ref{eq:marcus_rate} includes two key terms -- an Arrhenius-type contribution that is associated with free energy of solvent reorganization to bring reactant and product diabats into degeneracy and a prefactor that depends on the coupling between the diabatic states. 
RPMD
captures the solvent reorganization energetics because the path-integral-based method preserves exact quantum statistics. \cite{Par84,DeR84}  
The RPMD rate also correctly accounts for the tunneling contribution to the ET reaction rate, which can likewise be attributed to the path-integral basis of the method;  
the tunnel splitting for the electron between degenerate redox sites is analytically related to the reversible work for forming a kink-pair in the ring-polymer configuration.\cite{Cep95, Cha81,Mar91}
Given that the ensemble of reactive RPMD trajectories exhibit the dual rare events of solvent reorganization and kink-pair formation, and given that the FE barriers associated with these two steps are analytically related to the key terms in the Marcus rate expression, it is reasonable that Fig.~\ref{fig:atom_normal}(a) finds good agreement between RMPD and Marcus theory.
The RPMD method succeeds in the normal and activationless regimes because it captures the correct physics of the ET reaction.


Fig.~\ref{fig:atom_inverted} demonstrates that the success of the RPMD method does not extend into the inverted regime for ET, with both the RPMD rates and reaction mechanism deviating from the predictions of Marcus theory. 
  In Fig.~\ref{fig:atom_inverted}(a), the RPMD rates are seen to be only weakly dependent on the increasing driving force, rather than exhibiting the characteristic turnover in this inverted regime.  The RPMD trajectories also deviate from the reaction mechanism that is assumed in the Marcus TST, as is seen in Fig.~\ref{fig:atom_inverted}(b).  The reactive trajectories exhibit kink-pair formation directly from solvent configurations that are characteristic of the reactant basin; the expected solvent reorganization to configurations for which the electronic diabats are degenerate (indicated by the dashed line in the figure) is not observed.

To further explore the successes and failures of RPMD 
in these various regimes for ET, we compare the method with semiclassical instanton theory and exact quantum dynamics in the following section.

 \subsection{System-Bath Simulations}

In this section, we employ system-bath representations for ET to allow for the 
comparison of RPMD with other simulation techniques, 
including semiclassical instanton and exact quantum dynamics methods. 

\begin{figure}
\includegraphics{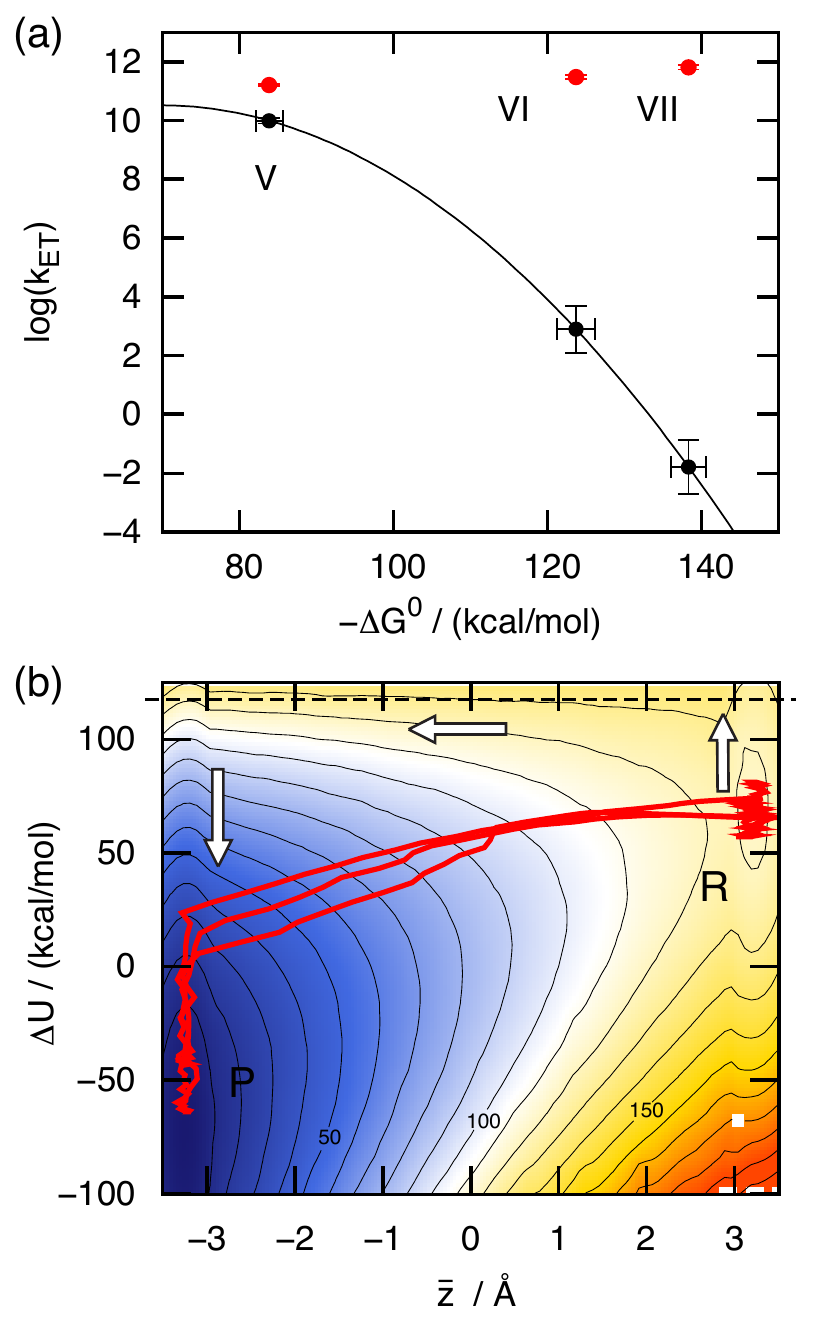}
\caption{(a)
ET reaction rates for the atomistic representation in the inverted regime, computed
using RPMD (red) and Marcus theory (black).  The various cases for the thermodynamic
driving force are labeled.
(b)  Representative trajectories (red) from the ensemble of reactive RPMD
trajectories for inverted ET (Case VI).  The trajectories are plotted as a function
of the ring-polymer centroid, $\bar{z}$, and the solvent collective variable,
$\Delta U$. The FE profile in these collective variables is also presented, with
contour lines indicating FE increments of 10 kcal/mol.
The white arrows indicate the solvent reorganization mechanism for ET that is
anticipated in the Marcus rate theory, and the dashed line indicates the value of
$\Delta U$ at which the reactant and product diabats cross in
Fig.~\ref{fig:atom_fes}(a).
}
\label{fig:atom_inverted}
\end{figure}

\begin{figure}
\includegraphics{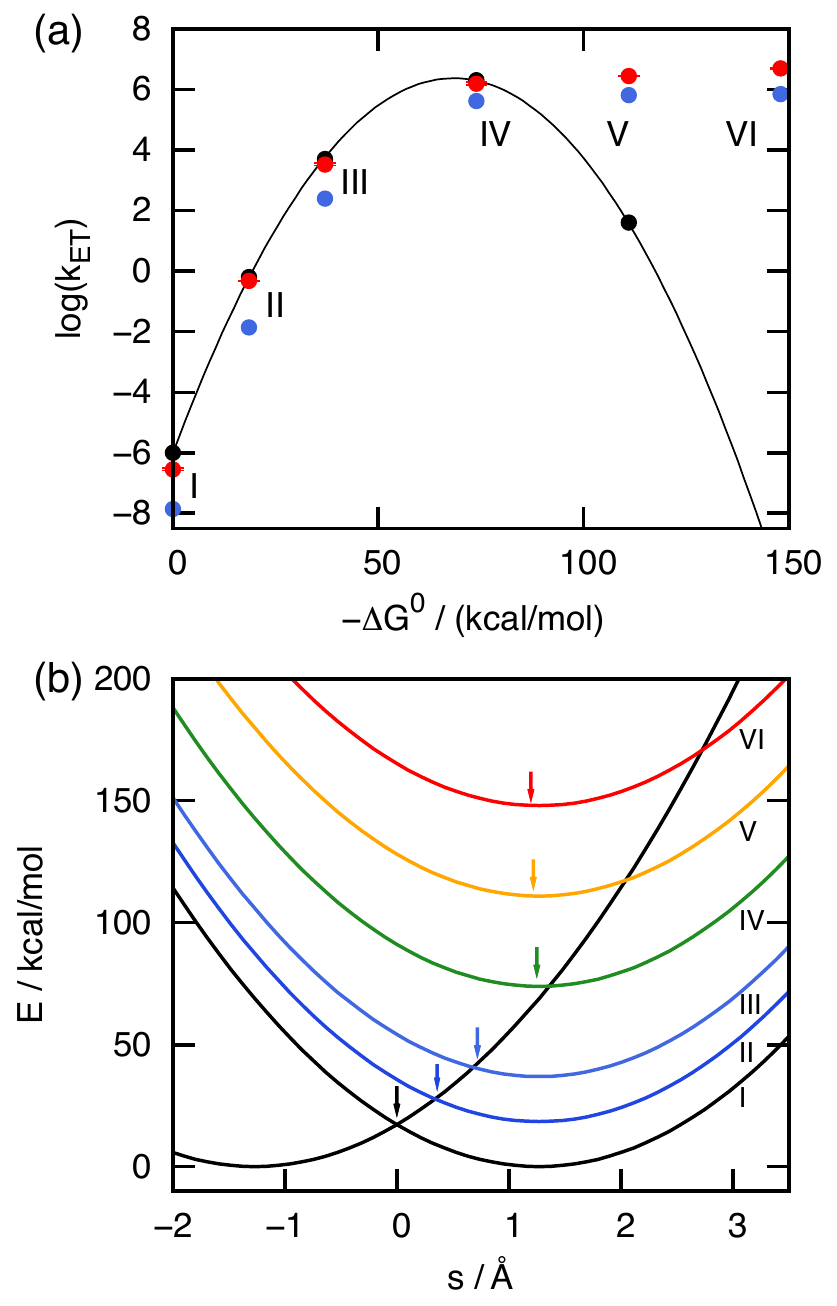}
\caption{(a) 
ET reaction rates for Model SB1, computed using RPMD (red), Marcus theory (black),
and SCI theory (blue).  
(b) FE profiles, $F (\Delta U)$,  for the reactant (colored, at right) and product
(left) diabatic electronic states as a function of the solvent coordinate, $s$. The
various cases of the thermodynamic driving force for the ET reaction are labeled;
see Table \ref{tab:model_asym_values} for details. 
The arrow indicates the value of the solvent coordinate that maximizes $\mathcal{I}_n (s)$, %
which corresponds to the dominant contribution to the SCI rate in Eq. \ref{eq:sci_rate_sd}.}
\label{fig:sb1_rate}
\end{figure}

Fig.~\ref{fig:sb1_rate}(a) and Table \ref{tab:rates_sb1} present a comparison of the RPMD and Marcus rates for Model SB1, which is parameterized to match the energy-scales  for the atomistic representation (Sec. \ref{sec:reduced_model}).  As before, the RPMD method reproduces the Marcus rates throughout the normal and activationless regimes,  while failing to predict the turnover of the ET rate in the inverted regime.
Analysis of the RPMD reactive trajectories in this system reveals mechanisms that are entirely analogous to those observed in Figs.~\ref{fig:atom_normal}(b), \ref{fig:atom_normal}(c), and \ref{fig:atom_inverted}(b) for the atomistic system.  Specifically, for the normal and activationless regimes, the RPMD trajectories exhibit
solvent reorganization to configurations for which the electronic diabats are degenerate, followed by rapid transfer of the electron between redox sites; 
and for the inverted regime, RPMD predicts ET without prior solvent reorganization.  These data confirm that Model SB1 exhibits the same essential physics as the atomistic representation.

\begin{table}
\caption{ET reaction rates for Model SB1, obtained using RPMD, Marcus theory, and SCI theory.\footnote{ET rates are given in $s^{-1}$, and the $-\Delta G^0$ are given in kcal/mol.}  \label{tab:rates_sb1} }
\begin{ruledtabular}
\begin{tabular}{crrrrrrrr}
 Case  & $-\Delta G^0$ & $ \log k_{\mathrm{MT}}$ & $\log k_{\mathrm{RPMD}}$ & $ \log k_{\mathrm{SCI}}$  & $\log \alpha k_{\mathrm{RPMD}}$ \\
\hline
I  &$0.0$     & $ -6.0$&  $ -6.55(4)$ & $-7.9 $ &   $-7.3$ \\
II & $18.5$  &$ -0.2$& $ -0.33(3)$ & $-1.9$ &   $-0.9$\\
III &$36.9$  & $  3.7$& $ 3.52(8)$ & $2.4$& $2.8$   \\
IV & $73.9$ & $  6.3$&  $ 6.19(5)$ & $5.6$& $5.4$\\
V &$110.9$ & $1.6$& $ 6.44(1)$ & $5.8$&  $5.8$ \\
VI  &$148.0$ &$-10.4$&  $ 6.69(3)$  &$5.9$& $5.9$  \\
\end{tabular}
\end{ruledtabular}
\end{table}

ET rates from the steepest-descent SCI theory (Eq.~\ref{eq:sci_rate_sd}) are also included in Fig.~\ref{fig:sb1_rate}(a) and Table \ref{tab:rates_sb1}.  Throughout the full range of thermodynamic driving forces, the instanton method tracks the RPMD results, 
including deviation from the Marcus predictions 
in the inverted regime.  
As is shown in Table \ref{tab:rates_sb1}, $\alpha$-correction of the RMPD rates (Eq. \ref{eq:alpha_correction}, assuming $\kappa_\mathrm{o}\approx1$) further improves their agreement with the SCI rates. 
These results underscore that the failure of RPMD does not arise from a breakdown in its formal connection with SCI theory;\cite{Ric09} instead, the comparison suggests that both RPMD and the SCI theory share the same underlying flaw in the inverted regime.
\footnote{Additional calculations performed using the RAW formulation of SCI theory\cite{Mil97,Mills_Chapter17} were found to be fully consistent with the SCI results in Fig.~~\ref{fig:sb1_rate}(a), but much more numerically unstable in the deep-tunneling regime considered here.}

The mechanistic predictions from SCI theory also show similarities with the RPMD results.  Fig.~\ref{fig:sb1_rate}(b)  presents the Marcus parabolas for the electronic diabats of Model SB1 as a function of the solvent coordinate, $s$.
Also shown are the solvent configurations that correspond to the SCI predictions for the ET transition state.
For each value of the thermodynamic driving force, the arrow in the figure indicates the solvent configuration that maximizes $\mathcal{I}_n(s)$, %
which corresponds to the largest contribution to the rate in Eq. \ref{eq:sci_rate_sd}.
For the normal and activationless regimes, SCI theory correctly predicts an ET transition state at the crossing of the electronic diabats.   However, in the inverted regime, the SCI transition state is instead located at the minimum of the reactant basin.
These mechanistic results from SCI theory are consistent with the observed pathways for the RPMD trajectories, which suggests that in the inverted regime, both RPMD and SCI theory overestimate the degree of ET from solvent configurations in the reactant basin.

To further illustrate this issue, we present SCI rate calculations for
deep tunneling in a 1D asymmetric double well.  Table
\ref{tab:rates_raw_dw} presents ET reaction rates calculated on the
potential energy surface $U_{\text{e-M}}(q)$ (Eq.~\ref{eq:u_redox}), with
parameters from Model SB1. Although this is a non-dissipative 1D system, the SCI rate is still
well-defined, and it is reported as a function of the potential energy
asymmetry.  The rates plateau to a finite value with increasing asymmetry,
which is consistent with rates for deep tunneling between a bound state
and a continuum.\cite{Jan99,Jan00,Hon90} However, this behavior is
qualitatively incorrect for tunneling rates between bound states, which
should vanish for non-degenerate states in accord with Fermi's Golden
Rule.\cite{Sakurai}
We conclude that  SCI theory, as well as
the closely related RPMD method, significantly overestimate the tunneling
probability between asymmetric bound states, leading to an incorrect ET
mechanism and overestimation of the reaction rate in the inverted regime.

\begin{table}
\caption{ET reaction rates for a 1D asymmetric double well, obtained using
SCI theory.\footnote{The Golden Rule for the symmetric case yields $\log k=-11.55$. $\Delta E$ is the difference between the two lowest eigenenergies for the system. All quantities reported in atomic units.}  \label{tab:rates_raw_dw} }
\begin{ruledtabular}
\begin{tabular}{ccc}
\multicolumn{1}{c}{$\epsilon$} & $\Delta E$ &\multicolumn{1}{c}{  $\log k_{\mathrm{SCI}}$}  \\
\hline
0.0 & 0.0 & $-11.0$\\
 0.05  & 0.02940 &$-10.9$\\
 0.10 & 0.05884 &$-10.8$\\
 0.20 & 0.11776 &$-10.6$\\
 0.30 & 0.17676 &$-10.3$\\
0.40 & 0.23584 &$-10.3$\\
\end{tabular}
\end{ruledtabular}
\end{table}

The results for the simple double-well system 
can be used to deduce a more general argument for the applicability of the RPMD and SCI calculations in ET problems.  Table \ref{tab:rates_raw_dw}, combined with the condition of detailed balance for the thermal reaction rate, indicates that the SCI 
rate for transfer in an asymmetric double-well system is approximately
\begin{equation}
k 
\approx\frac{2\pi}{\hbar}
|V_{12}|^2\ 
\text{min}\left(1,e^{-\beta\Delta E}\right).
\label{eq:rate_asym}
\end{equation}
For the Marcus-type ET mechanism in which electron tunneling is gated via solvent reorganization that symmetrizes the double-well system, Eq.~\ref{eq:rate_asym} leads to the TST rate in Eq.~\ref{eq:marcus_rate}.
However, for an unphysical  ``direct" ET mechanism in which electron tunneling proceeds from solvent configurations in the reactant basin (i.e., without prior solvent reorganization), Eq.~\ref{eq:rate_asym} leads to the following TST expression for the ET rate,
\begin{equation}
k_\mathrm{direct}=\frac{2\pi}{\hbar}
|V_{12}|^2 \left( \frac{\beta}{4\pi\lambda} \right)^\half
\text{min}\left(1,e^{-\beta\left(\lambda+\Delta G^0\right)}\right).
\label{eq:rate_direct}
\end{equation}

\begin{figure}
\includegraphics{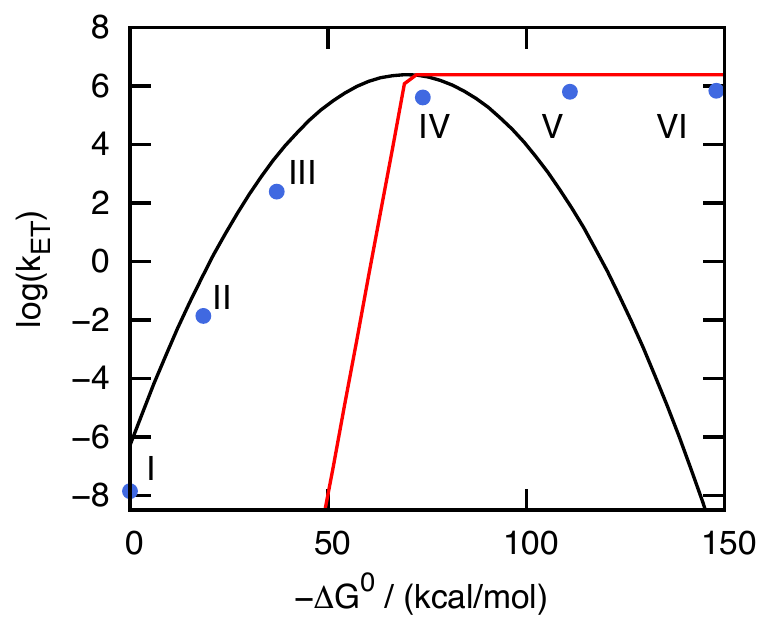}
\caption{The ET rates for Model SB1 corresponding to a
Marcus-like mechanism (black) and the ``direct" mechanism in Eq. \ref{eq:rate_direct} (red). SCI rates (blue) correspond to the kinetically favorable mechanism in all regimes.  See text for details.}
\label{fig:asym_tun}
\end{figure}

Fig.~\ref{fig:asym_tun} presents the ET reaction rates for Model SB1, 
assuming either the Marcus-type mechanism (Eq.~\ref{eq:marcus_rate}, black) or the direct mechanism (Eq.~\ref{eq:rate_direct}, red).  Also plotted are the rates calculated using SCI theory (Eq.~\ref{eq:sci_rate_sd}, blue).
Throughout the normal and activationless regimes, 
the rate for the Marcus-type mechanism dominates; in the inverted regime, the rate for the direct mechanism dominates; and the results from SCI theory closely track the larger of these two rates.
 It is clear that SCI theory (as well as RPMD) features a competition between the correct, Marcus-type mechanism for ET and the unphysical, direct mechanism for ET, and the prevailing mechanism is that which is predicted to be faster.
This analysis is fully consistent with the earlier discussions of the RPMD trajectories  (Figs.~\ref{fig:atom_normal}b, \ref{fig:atom_normal}c, and \ref{fig:atom_inverted}b) and the SCI transition state configurations (Fig.~\ref{fig:sb1_rate}b) for ET in the various regimes.
Furthermore, this analysis provides a general basis for expecting the SCI and RPMD methods to accurately describe ET rates in the normal and activationless regimes, and for expecting these methods to significantly overestimate the ET rate in the inverted regime.

\begin{figure}
\includegraphics{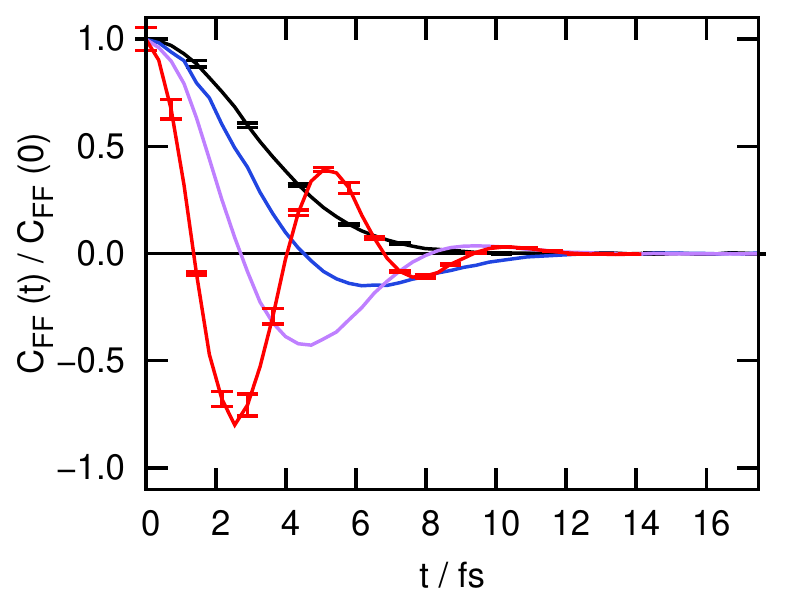}
\caption{Normalized flux-flux autocorrelation functions $C_{\mathrm{FF}}(t)$ for
Model SB2, calculated using exact quantum dynamics for Cases I (black), II (blue),
III (purple) and IV (red).
}
\label{fig:cfs}
\end{figure}

Table~\ref{tab:rates_sb2} presents ET rates for Model SB2, including results obtained using the QUAPI exact quantum dynamics method.  
Comparison of the RPMD, Marcus theory, and SCI theory rates for ET in Table~\ref{tab:rates_sb2} confirms that Model SB2 exhibits all of the previously discussed trends for these approximate methods.
Fig.~\ref{fig:cfs} presents the flux-flux correlation functions used to obtain the exact quantum rates for Model SB2.

%

\begin{table}
\caption{
ET reaction rates for Model SB2, obtained using RPMD, Marcus theory,  SCI theory, and exact quantum dynamics.\footnote{ ET rates are given in $s^{-1}$, and the $-\Delta G^0$ are given in kcal/mol.}  \label{tab:rates_sb2}}
\begin{ruledtabular}
\begin{tabular}{crrrrrrrrr}
 Case  & $-\Delta G^0$ & $ \log k_{\mathrm{MT}}$ & $\log k_{\mathrm{RPMD}}$ & $ \log k_{\mathrm{SCI}}$  & $ \log  k_{\mathrm{Q}}$ \\
\hline
I  &$0.0$        & $6.7$ &  $ 6.05(3)$ & $5.1$ & $6.7(1)$ \\
II & $5.3$   &$ 8.3$ & $7.73(5)$ & $6.6$ &      $8.5(1)$   \\
III &$8.8$   & $ 9.1$ & $ 8.54(3)$ & $7.4$    &   $    9.0 (3)$     \\
IV &  $17.6$ & $  9.8$ &  $ 9.27(2)$ & $8.6$   & $10.8(9)$    \\
V &$26.5$    & $8.9$  & $ 9.40(3)$  & $8.8$      &  $-$         \\
VI &$35.3$     &$6.3 $    &  $ 9.52(2)$  &$8.8 $         &   $-$     \\
\end{tabular}
\end{ruledtabular}
\end{table}

The results in Fig.~\ref{fig:cfs} emphasize the role of electronic state quantization in the ET reaction dynamics. 
At larger thermodynamic driving forces, the correlation functions become increasingly oscillatory, with a resonance frequency that matches the electronic state energy gap between the ET reactant and product.\cite{Ego04,Ego03} Integration over this increasingly oscillatory time correlation function (Eq.~\ref{eq:kquant}) contributes to the turnover in the ET reaction rate in the inverted regime.
The RPMD approximation to the real-time dynamics of the system, which is not expected to capture coherent quantum effects,\cite{Cra04,tfm08} does not fully enforce the quantization of electronic dynamics and leads to the observed inaccuracies in the inverted regime.  
Approximate quantum dynamical methods that 
explicitly enforce electronic quantization by using either a discrete electronic state basis or by exactly mapping to a continuous electronic basis
are thus expected to provide a better starting point for describing  state-to-state electronic dynamics and 
ET in the inverted regime.\cite{Kla97,Mey79,Tul98,Sto97,nandinitfm} 
Further investigation of this point  is in progress.

\section{Conclusions}

\label{sec:conclusion}

The current paper demonstrates the applicability of RPMD for the direct simulation of ET reaction dynamics in complex systems.
Using both atomistic and system-bath representations for ET in a polar solvent, we compare RPMD results with those obtained using Marcus theory, semiclassical instanton theory, and exact quantum dynamics.
Throughout the normal and activationless regimes for ET, RPMD correctly predicts the ET reaction mechanism and quantitatively describes the ET reaction rate over 12 orders of magnitude, without 
invoking any prior mechanistic or transition state assumptions.  Analysis of the RPMD trajectories reveals that the accuracy of the method lies in its exact description of statistical fluctuations, with regard to both solvent reorganization and the formation of kink-pair configurations during the electron tunneling event. 
However, for ET in the inverted regime, both RPMD and SCI theory fail 
 to predict the turnover in the ET reaction rate with increasing thermodynamic driving force.  In this regime, 
both methods overestimate the probability of electronic tunneling from solvent configurations in the reactant basin, leading to an overestimation of the corresponding reaction rates.  Exact quantum dynamics calculations illustrate that the limitations of the RPMD  method in the inverted regime arise from the inadequate 
quantization of the real-time electronic-state dynamics; 
analogous breakdowns of the method have been identified in other applications to strongly coherent quantum systems, including low-dimensional quantum oscillators \cite{Cra04} 
and electron-scattering in dilute fluids.\cite{tfm08}

We conclude by emphasizing that 
the normal and activationless regimes encompass the vast majority ET reactions in biological and synthetic systems.\cite{marcusNobelLecture}
The results presented here thus constitute a significant success for the RPMD method, demonstrating that it allows for the robust, direct simulation of thermally activated ET in systems with over 1000 atoms, leading to the quantitative prediction of ET reaction rates and and the potential discovery and characterization of ET reaction mechanisms in complex systems.
A comparable demonstration using other approximate real-time quantum simulation methods has not, to our knowledge, been previously reported.
Having established both the applicability and limitations of RPMD for ET reactions dynamics, this work provides the foundation for future studies of ET and proton-coupled ET reactions in enzymes and other condensed-phase systems.

\section{Acknowledgements}

\label{sec:acknowledge}
This work was supported by the 
U.S.~Office of Naval Research (USONR) under Grant No.~N00014-10-1-0884
National Science Foundation (NSF) CAREER Award under Grant No.~CHE-1057112.
Computing resources were provided by the National Energy Research Scientific
Computing Center (NERSC) and the Oak Ridge Leadership Computing Facility (OLCF).
T.F.M.~acknowledges support from a Camille and Henry Dreyfus Foundation New Faculty
Award and an Alfred P.~Sloan Foundation Research Fellowship.

\appendix{

\section{System-Bath Potential Energy Parameters}
\label{app:parabolic_caps}

\begin{table}[h!]
\caption{Parameters for the left Coulombic well in the electron-ion potential energy function of Eq.~\ref{eq:u_redox}
for Model SB1. \footnote{Unless otherwise noted, parameters are given in atomic units}\label{tab:para_fits_et_1}}
\begin{ruledtabular}
\begin{tabular}{ccccccccccc}
Case & $a_{\mathrm{L}}$& $b_{\mathrm{L}}$ & $c_{\mathrm{L}}$  &$r^{\mathrm{in}}_{\mathrm{L}}$ &$r^{\mathrm{out}}_{\mathrm{L}}$\\
\hline
I  &   $0.164567$&$2.002721$&$3.683127$&$-4.062912$&$-8.106702$\\
II &   $0.164520$&$2.001856$&$3.675494$&$-4.062912$&$-8.104948$\\
III  & $0.164472$&$2.000989$&$3.667859$&$-4.062912$&$-8.103197$\\
IV&   $0.164377$&$1.999251$&$3.652576$&$-4.062912$&$-8.099709$\\
V &   $0.164280$&$1.997506$&$3.637280$&$-4.062912$&$-8.096237$\\
VI &  $0.164183$&$1.995754$&$3.621971$&$-4.062912$&$-8.092782$\\
\end{tabular}
\end{ruledtabular}
\end{table}

\begin{table}[h!]
\caption{Parameters for the right Coulombic well in the electron-ion potential energy function of Eq.~\ref{eq:u_redox}
for Model SB1. \footnote{Unless otherwise noted, parameters are given in atomic units}\label{tab:para_fits_et_2}}
\begin{ruledtabular}
\begin{tabular}{ccccccccccc}
Case & $a_{\mathrm{R}}$ & $b_{\mathrm{R}}$&$c_{\mathrm{R}}$&$r^{\mathrm{in}}_{\mathrm{R}}$ &$r^{\mathrm{out}}_{\mathrm{R}}$ \\
\hline
I  &  $0.164567$&$-2.002721$&$3.683127$&$4.062912$&$8.106702$ \\
II &   $0.167357$&$-2.036963$&$3.752141$&$4.062912$&$8.108432$\\
III  & $0.170147$&$-2.071204$&$3.821152$&$4.062912$&$8.110110$\\
IV&   $0.175726$&$-2.139680$&$3.959165$&$4.062912$&$8.113319$\\
V &   $0.181304$&$-2.208150$&$4.097166$&$4.062912$&$8.116346$\\
VI & $0.186882$&$-2.276615$&$4.235157$&$4.062912$&$8.119207$\\
\end{tabular}
\end{ruledtabular}
\end{table}

\begin{table}[h!]
\caption{Parameters for the left Coulombic well in the electron-ion potential energy function of Eq.~\ref{eq:u_redox}
for Model SB2. \footnote{Unless otherwise noted, parameters are given in atomic units} \label{tab:para_fits_makri_1}}
\begin{ruledtabular}
\begin{tabular}{cccccccccccc}
Case & $a_{\mathrm{L}}$& $b_{\mathrm{L}}$ & $c_{\mathrm{L}}$  &$r^{\mathrm{in}}_{\mathrm{L}}$ &$r^{\mathrm{out}}_{\mathrm{L}}$\\
\hline
I   &  $0.157480$&$1.596286$&$1.609286$&$-3.050000$&$-7.086414$\\
II  &  $0.157507$&$1.596671$&$1.612042$&$-3.050000$&$-7.087150$\\
III &  $0.157525$&$1.596927$&$1.613880$&$-3.050000$&$-7.087642$\\
IV &  $0.157569$&$1.597568$&$1.618474$&$-3.050000$&$-7.088871$\\
V &   $0.157613$&$1.598208$&$1.623065$&$-3.050000$&$-7.090102$\\
VI &  $0.157657$&$1.598848$&$1.627656$&$-3.050000$&$-7.091336$\\
\end{tabular}
\end{ruledtabular}
\end{table}

\begin{table}[h!]
\caption{Parameters for the right Coulombic well in the electron-ion potential energy function of Eq.~\ref{eq:u_redox}
for Model SB2. \footnote{Unless otherwise noted, parameters are given in atomic units} \label{tab:para_fits_makri_2}}
\begin{ruledtabular}
\begin{tabular}{cccccccccccc}
Case &  $a_{\mathrm{R}}$ & $b_{\mathrm{R}}$&$c_{\mathrm{R}}$&$r^{\mathrm{in}}_{\mathrm{R}}$ &$r^{\mathrm{out}}_{\mathrm{R}}$ \\
\hline
I   &  $0.157480$&$-1.596286$&$1.609286$&$3.050000$&$7.086414$ \\
II  &  $0.156666$&$-1.587919$&$1.598482$&$3.050000$&$7.085675$ \\
III & $0.156124$&$-1.582341$&$1.591280$&$3.050000$&$7.085179$\\
IV & $0.154767$&$-1.568396$&$1.573273$&$3.050000$&$7.083924$\\
V &   $0.153410$&$-1.554450$&$1.555265$&$3.050000$&$7.082650$\\
VI &  $0.152053$&$-1.540504$&$1.537256$&$3.050000$&$7.081357$\\
\end{tabular}
\end{ruledtabular}
\end{table}


\newpage

\section{Transformation to Diabatic Basis}
\label{app:diabatic_localization}
The QUAPI method is implemented in an electronic diabatic state representation of the ET reaction.
In this appendix, we describe the procedure for transforming the potential energy function for Model SB2 from a position basis for the electron (Eq.~\ref{eq:sysbath}) to a diabatic basis where the reactant and product electronic states are maximally localized on the donor and acceptor metal atoms. 

We begin by calculating the two lowest adiabatic electronic eigenstates ($\psi_0(q;s)$ and $\psi_1(q;s)$) and eigenenergies ($E_0(s)$ and $E_1(s)$) of the system Hamiltonian 
at fixed values values of the solvent coordinate in the range $-8~\text{a}_0\le s \le 8~\text{a}_0$.
For each value of $s$, the system Hamiltonian is diagonalized on a uniform DVR grid of $1024$ electron positions in the range $-25 ~\text{a}_0\le q \le 25 \text~{a}_0$. 

For each value of $s$, reactant and product electronic wavefunctions in the diabatic basis are obtained via rotation of the two lowest-energy adiabatic wavefunctions, using
\begin{equation}
\label{eq:l_phi}
\phi_\mathrm{R}(q; s)=\cos(\theta_s) \psi_0 (q; s)-\sin(\theta_s) \psi_1(q; s)
\end{equation}
and
\begin{equation}
\label{eq:r_phi}
\phi_\mathrm{P}(q; s)=\sin(\theta_s) \psi_0(q; s)+\cos(\theta_s) \psi_1(q; s),
\end{equation}
where 
 \begin{equation}
 \theta_s=\frac{1}{2}\arctan\left(\frac{ S_{10}+ S_{01}}{S_{11}-S_{00}}\right)
 \end{equation}
 and 
 $S_{\mu\nu}= \int _{-\infty}^{0} {\psi_{\mu}(q;s)}^* \psi_{\nu}(q;s) \,d q$.
 This choice of the rotation angle, $\theta_s$, maximizes $\int _{-\infty}^{0} \left| \phi_\mathrm{R} (q; s)\right|^2 d q$, 
 the probability that the reactant diabatic state is positioned on the donor ion.
 Maximization of the probability that  the  product diabatic state is positioned on the acceptor ion yields an identical
 choice for $ \theta_s$.
 
The corresponding potential energy matrix elements in the diabatic basis (Eq.~\ref{eq:tls_hamiltonian}) are thus
\begin{align}
V_{11}(s)&=E_0(s)\cos^2\theta_s +E_1(s)\sin^2\theta_s, \\
V_{22}(s)&=E_0(s)\sin^2\theta_s +E_1(s)\cos^2\theta_s, \\
V_{21}(s)&=V_{12}(s)=\left(E_0(s)-E_1(s)\right)\cos\theta_s \sin \theta_s.
\end{align}
The diagonal elements are found to be parabolic functions of $s$, and the off-diagonal element are found to be nearly constant with respect to $s$.
We fit $V_{11}(s)$ and $V_{22}(s)$ to second-order polynomials functions (Eqs.~\ref{eq:v11} and \ref{eq:v22}) and employ a constant value for $V_{12}$ that corresponds to the $s=0$ result.  The polynomial expansion coefficients for $V_{11}(s)$ and $V_{22}(s)$ are provided in Table~\ref{tab:tls_params}, and 
the constant value for $V_{12}$ is provided in Table~\ref{tab:model_asym_values}.

\begin{table}
\caption{The diagonal elements of the diabatic potential matrix $V_{11}(s)$ and
$V_{22}(s)$ in Eqs.~\ref{eq:v11} and \ref{eq:v22} for Model SB2. \label{tab:tls_params}}
\begin{ruledtabular}
\begin{tabular}{ccccccccc}
 Case &$a_1 \times 10^{3}$ & $b_1 \times10^{2}$ & $c_1$ & $a_2  \times 10^{3} $ & $b_2 \times10^{2}$& $c_2$ \\
\hline
I & $4.7722$ &  $ 1.1308   $ & $-2.1576$ & $4.7722 $ &  $ 1.1308   $ & $-2.1576$ \\
II & $4.7722 $ &  $ 1.1308   $ & $-2.1477$ & $4.7722 $ &  $ 1.1308   $ & $-2.1561$ \\
III & $4.7722 $ &  $ 1.1308   $ & $-2.1411$ & $4.7721 $ &  $ 1.1308   $ & $-2.1551$ \\
IV &    $4.7720 $ &  $ 1.1307   $ & $-2.1245$ & $4.7720 $ &  $ 1.1308   $ & $-2.1526$ \\
\end{tabular}
\end{ruledtabular}
\end{table}

}



\bibliographystyle{apsrev} 

\end{document}